\documentclass[12pt,fleqn,a4paper]{article}
\usepackage{epsfig,latexsym,graphics, subfigure, cite}
\usepackage{amssymb}
\usepackage{feynarts}

\newcommand{\be}{\begin{equation}}
\newcommand{\ee}{\end{equation}}
\newcommand{\ba}{\begin{eqnarray}}
\newcommand{\ea}{\end{eqnarray}}
\newcommand{\bd}{\begin{displaymath}}
\newcommand{\ed}{\end{displaymath}}

\makeatletter
\renewcommand\appendix{\par
  \setcounter{section}{0}%
  \setcounter{subsection}{0}%
  \renewcommand\thesection{\@Alph\c@section}
}  
\makeatother

\begin{document}

\vspace*{-2cm}

\begin{flushright} 
\small{MPP-2009-157}
\end{flushright}

\vspace{1cm}

{

\centering

{ \bf \Large NLO electroweak contributions to gluino pair \\[0.3cm]  
production  at hadron colliders}

\vspace{1.cm}

{ \large \sc E.~Mirabella}

\vspace{0.2cm}
{\it Max-Planck-Institut f\"ur Physik
     (Werner-Heisenberg-Institut) \\ 
F\"ohringer Ring 6, 
D-80805 M\"unchen, Germany }

}

\vspace{0.3cm}

\begin{abstract}
\noindent
We calculate the full $\mathcal{O}(\alpha_s^2\alpha)$ corrections 
to the process of gluino pair production at hadron colliders
in the framework of the real MSSM. We
show that these contributions can be neglected at the LHC 
performing a scan over a 
wide region of the parameter space. The impact of these 
corrections in the   
parameter range  investigated  at the Tevatron is small.
\end{abstract}

\vspace{0.3cm}

\section{Introduction}
Despite of its success, the Standard Model (SM) of particle physics is affected by theoretical
and phenomenological problems whose solution can be used as a guideline for its extension.
In this respect Supersymmetry (SUSY)~\cite{Wess:1974tw,Volkov:1973ix}, and in particular the Minimal Supersymmetric extension of the 
Standard Model (MSSM)~\cite{Nilles:1983ge,Haber:1984rc,Barbieri:1987xf},
is one of the most promising scenarios for physics beyond the Standard Model.
In SUSY  models the  breaking of electroweak symmetry is obtained radiatively 
at a scale which is stabilized by  Supersymmetry itself.  Moreover, the comparison of MSSM 
predictions and electroweak precision observables  provides an overall fit of data~\cite{Ellis:2007fu,Buchmueller:2007zk} which is 
at least as good as that obtained using the SM, and even better in the case of specific observables such as $g-2$ of the 
muon~\cite{Bennett:2002jb,Bennett:2004pv}.
\medskip

These features render SUSY an appealing framework, and 
they can explain the big effort in the hunting for SUSY not only in the past but also in  future years.
In particular, this search is one of the major goals of the LHC. Experimental studies 
have shown the possibility of  early discovery of SUSY   with $1~\mbox{fb}^{-1}$ of integrated luminosity in the inclusive multijet plus 
missing $E_T$ channel~\cite{Aad:2009wy,0954-3899-34-6-S01}, provided that the masses of squarks and gluinos are not too heavy ({\sl e.g.} $<2$~TeV). Interestingly, the 
$95~\%$ confidence level area of the $(m_0, m_{1/2})$ plane of the Constrained MSSM (CMSSM) lies largely in the region that will be investigated  with  $1~\mbox{fb}^{-1}$
at $14$~TeV~\cite{Buchmueller:2007zk}. 
\medskip

At hadron colliders, colored particles can be searched for most efficiently. This implies that
Supersymmetry could be discovered looking at the production of squarks and gluinos.
Among the others, the process of gluino pair production,
\begin{equation}
\label{Eq:ProcessGLU}
P~P \to \tilde{g}~\tilde{g}~X,
\end{equation}
is  one of the most important processes  leading to the
production of colored SUSY particles. Indeed,  
its  cross section is large, $\mathcal{O}(10\mbox{~pb})$  if the  gluino mass is $\mathcal{O}(600\mbox{~GeV})$. Moreover,
 the gluino plays a key role in characterizing SUSY models. The measure of  the spin of a (supposed to be)
gluino~\cite{Alves:2006df} and the confirmation of its  
Majorana nature~\cite{Barnett:1993ea,Kraml:2005kb,Choi:2008pi} would allow  not only  to distinguish  
among different beyond Standard Model scenarios,  but also among MSSM and  others SUSY models involving Dirac gauginos, such as the 
$N=1 / N=2$  hybrid scheme~\cite{:2008gva}.  
\medskip

The total cross section for gluino pair production was computed at Born level 
long time ago~\cite{Kane:1982hw,Harrison:1982yi,Reya:1984yz,Dawson:1983fw}.  
NLO SUSY-QCD contributions  were computed  in Ref.~\cite{Beenakker:1996ch}. These corrections  
are  positive and large 
\mbox{(from} $5$ to $90\%$, depending on the masses of the squarks and of the gluino),
and they reduce appreciably the factorization scale dependence. They are included into the 
publicly available code \verb|Prospino|~\cite{Beenakker:1996ed}. 
More recent is the resummation of the
QCD Sudakov logarithms at the
next-to-leading-logarithmic (NLL) accuracy~\cite{Kulesza:2008jb, Kulesza:2009kq}, and
the resummation of the leading Coulomb corrections~\cite{Kulesza:2009kq}. Their inclusion further stabilizes
the prediction against scale variation. The NLL contributions are of the order of
$2-8 \%$ of the NLO QCD predictions, provided the squark and gluino
masses are $\mathcal{O}(1~\mbox{TeV})$. In this mass range the contribution of the
Coulomb corrections amounts up to $5 \%$.  
The computation of NLO electroweak (EW) corrections, of $\mathcal{O}(\alpha_s^2 \alpha)$, to the process of hadronic production of a gluino pair is still missing.
In this paper, we fill this gap computing the full $\mathcal{O}(\alpha_s^2 \alpha)$ corrections 
to the process~(\ref{Eq:ProcessGLU})  
in the framework of the real MSSM. Our computation is part of an ongoing project
aiming to evaluate the tree-level EW and NLO EW contributions to the production of colored SUSY particles at 
the LHC~\cite{Hollik:2007wf, Beccaria:2008mi, Bornhauser:2007bf, Hollik:2008yi, Hollik:2008vm}. 
\medskip

The plan of the paper is the following. In section~\ref{Sec:3levelGLU} we 
briefly summarize the $\mathcal{O}(\alpha_s^2)$ contributions to the process~(\ref{Eq:ProcessGLU}). In 
section~\ref{S:ObigAlfaGLU} we describe the partonic processes contributing at  $\mathcal{O}(\alpha_s^2 \alpha)$. Numerical 
results for the electroweak corrections to gluino pair 
production at the LHC are presented in section~\ref{Sec:NumericsLHC}, while section~\ref{Sec:NumericsTEV} is devoted to
a brief discussion on the numerical value of the electroweak
corrections at the Tevatron. Section~\ref{Conclusions} 
summarizes our results. A list of Feynman diagrams is  collected in the 
appendix.


\section{Gluino pair production in lowest order}
\label{Sec:3levelGLU}
The leading order contributions to the process
(\ref{Eq:ProcessGLU}) are of QCD origin, of $\mathcal{O}(\alpha_s^2)$.
At lowest order in the perturbative expansion the differential  cross
section  can be written as follows,
\begin{eqnarray}
\label{Eq:DiffXSecGLU}
d\sigma^{\mbox{\tiny QCD, LO}}_{PP\to \tilde{g}\tilde{g}}(S) &=& \sum_{q} \int_{\tau_0}^{1}~d\tau \frac{dL_{q\overline{q}}}{d\tau}
(\tau)d \sigma^{2,0}_{q\overline{q}\to \tilde{g}\tilde{g}}(\tau S) + 
\int_{\tau_0}^{1}~d\tau \frac{dL_{gg}}{d\tau}
(\tau)d \sigma^{2,0}_{gg\to \tilde{g}\tilde{g}}(\tau S), \nonumber \\
\end{eqnarray}
with the help of the parton luminosities, defined according to
\begin{equation}
\label{Eq:Lumi}
\frac{dL_{ij}}{d\tau}(\tau)= \frac{1}{1+\delta_{ij}}~
\int_{\tau}^1 \frac{dx}{x} \left [ f_{i|P}(x)f_{j|P}\left(\frac{\tau}{x}\right)+ 
f_{j|P}(x)f_{i|P}\left(\frac{\tau}{x}\right) \right ] .
\end{equation}
$f_{i|H}(x)$ is the momentum distribution of the parton $i$ inside the hadron $H$.
The sum runs over the quarks $q=u,~d,~c,~s,~b$.  
The lower limit on the integral over $\tau$, 
$\tau_0= 4\, m^2_{\tilde{g}}/ S$, is related to
the threshold for the production of the gluino pair.
We use the convention $d \sigma^{a,b}_{X}$ to denote the cross section for a partonic process $X$
at a given order $\mathcal{O}(\alpha_s^a \alpha^b)$ in the strong and electroweak coupling constants.
Therefore,
$d\sigma^{2,0}_{q \bar{q} \to\tilde{g}\tilde{g}}$ and $~d\sigma^{2,0}_{gg\to\tilde{g}\tilde{g}}$  
are the lowest order differential  cross sections for the partonic processes
\begin{eqnarray}
\label{Eq:PartProGLUQ}
q(p_1)~\overline{q}(p_2) &\to& ~\tilde{g}(k_1)~\tilde{g}(k_2), \\
\label{Eq:PartProGLUG}
g(p_1)~g(p_2) &\to& ~\tilde{g}(k_1)~\tilde{g}(k_2), 
\end{eqnarray}
respectively. 
The cross sections are averaged (summed) over the spins and the 
colors of the incoming (outgoing) particles.  In this analysis we will consider the five light quarks as massless and we approximate the 
CKM matrix by  the unity matrix. We 
retain the mass of the bottom in the Yukawa couplings, owing to the possible enhancement due to $\tan \beta$.  We perform our 
computation in Feynman gauge. 
\medskip

In lowest order, 
the partonic cross sections for the processes~(\ref{Eq:PartProGLUQ}) and (\ref{Eq:PartProGLUG})
can be obtained from the Feynman diagrams in Fig.~\ref{Fig:TREE} of
appendix~\ref{SSec:FeynmanDiagramsGLUGLU}. The cross sections 
can be written as
\begin{eqnarray}
d \sigma^{2,0}_{q \bar{q}  \to \tilde{g} \tilde{g}} &=& 
\frac{dt}{16 \pi s^2} \; \overline{\sum} \left | \mathcal{M}^{1,0}_{q
    \bar{q} \to \tilde{g}\tilde{g}}\right |^2, \nonumber \\[0.6ex]
d \sigma^{2,0}_{g g  \to \tilde{g} \tilde{g}} &=&  
\frac{dt}{16 \pi s^2} \; \overline{\sum} \left | \mathcal{M}^{1,0}_{gg\to \tilde{g}\tilde{g}}\right |^2, 
\end{eqnarray}
where $\mathcal{M}^{1,0}_{X}$ is the tree-level contribution to the
amplitude of the process $X$. The squared amplitudes averaged (summed) over the spins and the colors of
the initial (final) particles 
read~\cite{Kane:1982hw,Harrison:1982yi,Reya:1984yz,Dawson:1983fw,Beenakker:1996ch}
\begin{eqnarray}
\overline{\sum} \left | \mathcal{M}^{1,0}_{q
    \bar{q} \to \tilde{g}\tilde{g}}\right |^2 &=& \frac{8}{27} \, \alpha_s^2 \,
\pi^2 \Bigg \{ \frac{72}{s^2}\left(2m^2_{\tilde{g}}s+t_{\tilde{g}}^2+u_{\tilde g}^2
\right) + 4 m^2_{\tilde{g}} s \left (   \frac{1}{u_{\tilde{q},1}t_{\tilde{q},1}} + \frac{1}{u_{\tilde{q},2}t_{\tilde{q},2}}\right) 
\nonumber \\
&+& \frac{36(m^2_{\tilde{g}} s +t_{\tilde g}^2)}{s} \left(
  \frac{1}{t_{\tilde{q},1}} + \frac{1}{t_{\tilde{q},2}}\right)
+ 16 t^2_{\tilde g} \left(
  \frac{1}{t^2_{\tilde{q},1}} +
  \frac{1}{t^2_{\tilde{q},2}}\right) 
\nonumber \\
&+&  \frac{36(m^2_{\tilde{g}} s +u_{\tilde g}^2)}{s} \left(
  \frac{1}{u_{\tilde{q},1}} + \frac{1}{u_{\tilde{q},2}}\right) 
+  16 u^2_{\tilde g} \left(
  \frac{1}{u^2_{\tilde{q},1}} +
  \frac{1}{u^2_{\tilde{q},2}}\right) \Bigg \},  \nonumber
 \\[0.6ex]
\overline{\sum} \left | \mathcal{M}^{1,0}_{gg\to \tilde{g}\tilde{g}}\right |^2 
 &=& 18  \,  \alpha_s^2 \, \pi^2   \Bigg \{  \left ( 1 - \frac{t_{\tilde g} u_{\tilde g}}{s^2}
\right ) \left [ \frac{s^2}{t_{\tilde g} u_{\tilde g}} -2 + 4
  \frac{m^2_{\tilde{g}}s}{t_{\tilde g} u_{\tilde g}} \left (1
    -\frac{m^2_{\tilde{g}}s}{t_{\tilde g} u_{\tilde g}} \right ) \right ] \Bigg \}. 
\end{eqnarray}
A factor $1/2$ has been taken into account  because of the identical particles in the final states. 
The Mandelstam variables are defined as
\begin{eqnarray}
\label{Eq:Mandel}
s &=& (p_1 + p_2)^2,  \\
t &=& (p_2 - k_2)^2,~~~~ t_{\tilde{q},a} = t - m^2_{\tilde{q},a},~~~~ t_{\tilde g} = t-m^2_{\tilde{g}}, \nonumber \\
u &=& (p_1 - k_2)^2,~~~~ u_{\tilde{q},a} = u - m^2_{\tilde{q},a},~~~~ u_{\tilde g} = u-m^2_{\tilde{g}}, \nonumber 
\end{eqnarray}
and $u = 2 m^2_{\tilde{g}} - s - t$.


\section{\boldmath{$\mathcal{O}(\alpha_s^2\alpha)$} corrections to the hadronic process}
\label{S:ObigAlfaGLU}
Gluinos do not interact weakly, thus  a pair of gluinos is neither  produced  at 
$\mathcal{O}(\alpha^2)$, from $q \bar{q}$ initial states, 
nor at $\mathcal{O}(\alpha_s\alpha)$ via photon-induced processes. Therefore, in contrast to 
squark--anti-squark~\cite{Hollik:2007wf,Bornhauser:2007bf,Hollik:2008yi} and squark-gluino~\cite{Hollik:2008vm} production, 
EW contributions enter only at NLO, and they are at least of $\mathcal{O}(\alpha_s^2 \alpha)$. 
The NLO EW contributions to the hadronic differential cross section reads as follows,
\begin{eqnarray}
\label{Eq:MainFormulaGLU}
 d\sigma^{\mbox{\tiny EW NLO}}_{PP\to\tilde{g}\tilde{g}X}(S) &=& \sum_{q} \Bigg  \{ \int_{\tau_0}^1 d\tau \Bigg[ 
\frac{dL_{q \overline{q}}}{d\tau}(\tau) 
    \bigg ( d\sigma^{2,1}_{q \overline{q} \to \tilde{g}\tilde{g}}(\tau S)
    +d\sigma^{2,1}_{q \overline{q} \to \tilde{g}\tilde{g}\gamma}(\tau S)
    \bigg)  
\nonumber \\
&+& \frac{dL_{q \gamma }}{d\tau}(\tau) d\sigma^{2,1}_{q \gamma \to
  \tilde{g}\tilde{g} q}(\tau S) +                                       
\frac{dL_{\gamma \overline{q}}}{d\tau}(\tau) d\sigma^{2,1}_{\gamma
  \overline{q} \to \tilde{g}\tilde{g}\overline{q}}(\tau S)   \Bigg] \Bigg \}.
\end{eqnarray}
The $q \gamma$ and $\gamma\bar{q}$ luminosities entering~(\ref{Eq:MainFormulaGLU}) 
are built according to Eq.~(\ref{Eq:Lumi}). Besides the virtual corrections and the real photon radiation 
processes at $\mathcal{O}(\alpha_s^2 \alpha)$, we consider the photon-induced processes
leading to the production of a gluino pair together with an (anti-)quark. 
Diagrams and amplitudes are generated with
\verb|FeynArts|~\cite{Hahn:2000kx,Hahn:2001rv}. The reduction  of  the 
one-loop integrals is performed with the help of
\verb|FormCalc|~\cite{Hahn:1998yk,Hahn:2006qw}, while 
the scalar one-loop integrals are numerically evaluated using
\verb|LoopTools|.
Infrared (IR) and collinear singularities are treated using mass regularization, 
{\sl i.e.} giving a small mass to the photon and to the five light quarks.


\subsection{\boldmath{$q \bar{q}$} annihilation with electroweak loops}
\label{S:loopCorrection}
The first class of corrections entering Eq.~(\ref{Eq:MainFormulaGLU})
are the electroweak one-loop corrections 
to the processes (\ref{Eq:PartProGLUQ}), yielding the following partonic
cross section,
\begin{equation}
d \sigma^{2,1}_{q \overline{q}\to \tilde{g}\tilde{g}} = 
 \frac{dt}{16 \pi s^2} \overline{\sum}  2 \, \mathfrak{Re} \left \{ \mathcal{M}^{1,0*}_{q\overline{q}\to \tilde{g}\tilde{g}}
\mathcal{M}^{1,1}_{q\overline{q}\to \tilde{g}\tilde{g}}  \right \}.
\end{equation}
$\mathcal{M}^{1,1}_{q\overline{q}\to \tilde{g}\tilde{g}}$ is the
one-loop electroweak
contribution to the amplitude
of the quark--anti-quark annihilation process. 
The diagrams responsible for this contribution  are displayed in
 Fig.~\ref{Fig:LOOPQ} of appendix~\ref{SSec:FeynmanDiagramsGLUGLU}.  \\
\noindent

We treat UV divergences using dimensional reduction. 
In order to cure the UV divergences we have to renormalize the quark
and the squark sector at $\mathcal{O}(\alpha)$. Renormalization of mass and wavefunction of the
quarks and squarks belonging to the first two generations, 
of $\tan \beta$ and of the mass of the $W$ boson, has been performed according to the procedure described in Ref.~\cite{Hollik:2008yi}. The 
renormalization of mass and wavefunction  of the bottom and top quarks and squarks has been 
widely studied~\cite{Eberl:2001eu,Brignole:2002bz,Oller:2003ge,Hollik:2003jj,Heinemeyer:2004xw}, and several 
renormalization schemes have been proposed. Each of these schemes has 
its own virtues and drawbacks, we perform our computation using two
different renormalization schemes. The first (second) scheme, referred
in the following as Rs1 (Rs2), is the "$m_b$ OS" ("$m_b$
$\overline{\mbox{DR}}$") 
scheme defined in Ref.~\cite{Heinemeyer:2004xw}. The
renormalization of the stop-sbottom sector at $\mathcal{O}(\alpha)$
within the Rs2 scheme requires the renormalization of the
supersymmetric Higgs parameter $\mu$. This parameters has been defined 
in the $\overline{\mbox{DR}}$ scheme.
\medskip

In the case of $b \bar{b} \to \tilde{g} \tilde{g}$ we keep the mass
of the b-quark  that appears in the couplings. In this case, the last twelve diagrams in
Fig.~\ref{Fig:LOOPQ}  of the appendix~\ref{SSec:FeynmanDiagramsGLUGLU}
have
to be considered as well. It is well
known~\cite{Banks:1987iu,Hall:1993gn,Hempfling:1993kv,Carena:1994bv,Carena:1999py,Eberl:1999he}
that, in the large
$\tan \beta$ regime, the tree-level relation between the bottom mass $m_b$
and the bottom Yukawa couplings $y_{b}$  receives radiative corrections that can be
strongly enhanced and have to be resummed. Power counting in $\alpha_s
\tan \beta$  shows that the leading $\tan \beta$ enhanced contributions,
of $\mathcal{O}(\alpha^n_s \tan^n \beta)$, can be
accounted for by means of the substitution
\begin{equation}
m^{\mbox{\tiny Rs}}_b \to \overline{m}^{\mbox{\tiny Rs}}_b =
\frac{m^{\mbox{\tiny Rs}}_b }{1+\Delta_b}
\end{equation}
in the relation between $m_b$ and $y_{b}$. 
$m^{\mbox{\tiny Rs}}_b$ is the bottom
mass in a given
renormalization scheme, Rs. $\Delta_b$ is defined as
\begin{eqnarray}
 \Delta_b &=&  \frac{2 \alpha_s}{3 \pi} \frac{m_{\tilde{g}} \; \mu \; \tan \beta}{(m^2_{\tilde{b},1}-m^2_{\tilde{b},2})(m^2_{\tilde{b},2}-m^2_{\tilde{g}})
(m^2_{\tilde{b},1}-m^2_{\tilde{g}})}\left [  m^2_{\tilde{b},1} m^2_{\tilde{b},2} \ln \left(\frac{m^2_{\tilde{b},1}}{m^2_{\tilde{b},2}}
  \right) \right. \nonumber \\
&+& \left. m^2_{\tilde{b},2} m^2_{\tilde{g}} \ln \left(\frac{m^2_{\tilde{b},2}}{m^2_{\tilde{g}}} \right)  
+ m^2_{\tilde{g}} m^2_{\tilde{b},1} \ln
  \left(\frac{m^2_{\tilde{g}}}{m^2_{\tilde{b},1}} \right)  \right ].
\end{eqnarray}
Concerning the Higgs sector, the $b-\bar{b}-h^0_u$ coupling is
dynamically generated at $\mathcal{O}(\alpha_s)$. This coupling can be enhanced 
if $\tan \beta$ is large and it is worth to include such effects
modifying the $b-\bar{b}-$Higgs Yukawa couplings. In particular, the
effective Lagrangian that correctly takes into account these
dynamically generated extra-couplings is
\begin{eqnarray}
\mathcal{L}^{\mbox{\tiny eff.}}_{\mbox{\tiny Higgs}} &=& 
\frac{\overline{m}^{\mbox{\tiny Rs}}_b}{v} \Bigg [ \tan \beta \left (
  1 - \frac{\Delta_b}{\tan^2 \beta}
\right ) A^0 \bar{b} i \gamma^5 b  + \frac{\sin \alpha}{\cos \beta} \left (
  1 - \frac{\Delta_b}{\tan \alpha \tan \beta}  
\right ) h^0 \bar{b} b  \nonumber \\
&~& - \frac{\cos \alpha}{\cos \beta} \left (1 +
  \frac{\Delta_b \tan \alpha}{\tan \beta}    \right )  H^0 \bar{b} b  \Bigg ].
\end{eqnarray}


\subsection{\boldmath{$q \bar{q}$} annihilation with real photon emission}
\label{Sec:Reale}
IR singularities in the virtual corrections are cancelled when the tree-level  contribution of the partonic process of real photon 
radiation, 
\begin{equation}
\label{Eq:PartonGamma}
q(p_1)~\overline{q}(p_2) \to ~\tilde{g}(k_1)~\tilde{g}(k_2)~\gamma(k_3), 
\end{equation}
is included. This contribution can be computed using the Feynman diagrams depicted
in Fig.~\ref{Fig:2in3gamma} of appendix~\ref{SSec:FeynmanDiagramsGLUGLU}. The integral over the phase space is IR divergent
when $k_3 \to 0$, while  collinear singularities appear whenever $k_3 \cdot p_i \to 0$.  IR and collinear divergences appearing in the phase space integration are regularized
using both, phase space slicing~\cite{Baier:1973ms,Denner:1991kt, Hollik:2007sq} and dipole subtraction~\cite{Dittmaier:1999mb,Dittmaier:2008md}. The two methods 
are in good numerical agreement, as found also in  the case of squark-anti--squark production~\cite{Hollik:2008yi}. As already mentioned, IR singularities 
cancel when the real radiation processes and the virtual contributions are added together, as in  Eq.~(\ref{Eq:MainFormulaGLU}). Collinear 
singularities remain and have to be absorbed via the factorization of the parton distribution functions (PDFs), {\sl c.f.} section~\ref{Sec:Facto}.


\subsection{\boldmath{$q \gamma$} and \boldmath{$\gamma \bar{q}$} fusion}
\label{SSec:RealEmi} 
The last class of  $\mathcal{O}(\alpha_s^2\alpha)$ contributions to the process (\ref{Eq:ProcessGLU})
are  the tree-level contributions of the partonic processes
\begin{eqnarray}
\label{Eq:gammaquark}
q(p_1)~      \gamma(p_2) &\to& ~\tilde{g}(k_1)~\tilde{g}(k_2)~          q (k_3), \\
\label{Eq:gammaantiquark}
\gamma(p_1)~\overline{q}(p_2) &\to& ~\tilde{g}(k_1)~\tilde{g}(k_2)~\overline{q}(k_3). 
\end{eqnarray}
These contributions can be computed from the Feynman diagrams
depicted in Fig.~\ref{Fig:QuarkGamma} of appendix~\ref{SSec:FeynmanDiagramsGLUGLU}. 
\medskip

Note that, if $m_{\tilde{q}} > m_{\tilde{g}}$, the quark in the final state
can be the decay product of an 
on-shell squark. If this is 
the case the last four diagrams depicted in Fig.~\ref{Fig:QuarkGamma}  become
singular. The related poles have to be regularized inserting the width of the on-shell
squarks into the corresponding  propagator. Furthermore, the contribution obtained squaring the resonant diagrams 
has to be subtracted since it arises from the production and the subsequent 
decay of an (anti-)squark through (anti-)quark--photon fusion,
\begin{eqnarray}
q~\gamma~\to~\tilde{g}\tilde{q}^{~} &\mbox{~~ and ~~}& \tilde{q}^{~}~\to~\tilde{g}~q, \nonumber \\
\gamma~\bar q~\to~\tilde{g}\tilde{q}^{*} &\mbox{~~ and ~~}& \tilde{q}^{*}~\to~\tilde{g}~\bar{q}. 
\end{eqnarray}
According to Refs.~\cite{Beenakker:1996ch,Hollik:2008vm}, the extraction of the Breit-Wigner
pole contribution has been performed in the narrow width approximation. 
\medskip

Collinear singularities arising from initial state emission are again absorbed into the PDFs. 
These singularities  are regularized  using 
both, phase space slicing and dipole subtraction. The formulae needed can be found in Ref.~\cite{Baier:1973ms}
and in Refs.~\cite{Diener:2005me, Dittmaier:2008md}, respectively. The
results obtained using the two methods agree within the integration uncertainty.  
\medskip

The contribution of this channel is expected to be small owing to the
suppression of the photon PDF inside the proton. Indeed, the photon
PDF is intrinsically suppressed with respect to the valence quark PDF
by a factor $\alpha$, since this PDF  is originated
from the emission  of a photon from  a  (anti-)quark. In the SUSY scenarios  we consider, the
contribution of this partonic process amounts up to few percent of the
whole $\mathcal{O}(\alpha_s^2 \alpha)$ correction.


\subsection{Factorization of initial collinear singularities}
\label{Sec:Facto}
As already mentioned,  the universal logarithmic divergences  related to  the collinear splittings
\begin{displaymath}
q~\to~q~\gamma,~~~~\bar{q}~\to~\bar{q}~\gamma,~~~\gamma~\to~q~\bar{q},
\end{displaymath}
are absorbed into the definition of the PDFs via mass factorization. We factorize the (anti-)quark PDFs at
$\mathcal{O}(\alpha)$ in the DIS scheme. 
The effect of this factorization  is to add the following term 
into Eq.~(\ref{Eq:MainFormulaGLU}),
\begin{eqnarray}
d \sigma^{\mbox{\tiny Fact.}}_{PP \to \tilde g \tilde gX}(S) 
&=& \int_{\tau_0}^1 \, d \tau \, \sum_{q}  \Bigg \{ \bigg [  -\frac{\alpha}{\pi} e^2_q\frac{dL_{q \overline{q}}}{d\tau}(\tau)
                                        \int_{z_0}^1 dz  
                            \left[\mathcal{H}^{(1)}_q \right                            ]_+ 
 d \sigma^{2,0}_{q\overline{q}\to \tilde{g}\tilde{g}}(z \tau S) \nonumber \\
 &-& \frac{3\alpha}{2 \pi} e^2_q\left ( \frac{dL_{q \gamma  }}{d\tau}(\tau) + \frac{dL_{\gamma
      \overline{q}}}{d\tau}(\tau) \right )                                         \int_{z_0}^1 dz  
                             ~ \left (\mathcal{H}^{(2)}_q  \right )   d \sigma^{2,0}_{q \overline{q}\to \tilde{g}\tilde{g}}(z \tau S)
\bigg ]
\Bigg                            \}. \nonumber \\
\label{Eq:erre}
\end{eqnarray}
The functions $\mathcal{H}_q^{(1)}$ and  $\mathcal{H}_q^{(2)}$ read as follows,
\begin{eqnarray}
\mathcal{H}_q^{(1)} &=& P_{qq}(z) \left [ \ln \left ( \frac{\mu^2_F}{m^2_q}
    \frac{1}{z(1-z)} \right ) -1 \right ] -\frac{3}{2}\frac{1}{1-z} + 2z+3, \nonumber \\
\mathcal{H}_q^{(2)} &=& P_{q\gamma}(z) \ln \left (
  \frac{\mu^2_F}{m^2_q}\frac{1-z}{z}\right )-1 + 8z - 8z^2,
\end{eqnarray}
where the splitting functions are 
\begin{displaymath}
P_{qq}(z) = \frac{1+z^2}{1-z},~~~~~ P_{q\gamma}(z) = z^2 + (1-z)^2. 
\end{displaymath}
$z_0$
is defined as $z_0 = 4 m_{\tilde{g}} / (\tau S)$, while $e_q$ is the charge of the quark $q$ 
expressed in  units of the positron charge.  The $\left [ \cdots \right ]_+$
distribution is defined as
\begin{equation}
\int_a^1 \; dx \; [f(x)]_+ g(x)  =  \int_a^1 \; dx \;  f(x) \big [
g(x) - g(1) \big ] - g(1) \; \int_0^a \; dx \; f(x).  
\label{Eq:PlusD}
\end{equation}
In the actual
computation, we use the MRST2004qed parton
distribution functions  at NLO~QED and 
NLO~QCD~\cite{Martin:2004dh}. This fit takes into
account  QED-effects into the DGLAP evolution equations and the
parametrization of the PDF at the initial scale. 
MRST2004qed PDFs are defined at NLO QCD within the
$\overline{\mbox{MS}}$ mass factorization scheme. As discussed in
Ref.~\cite{Diener:2005me}, the DIS scheme is used for the factorization of the
$\mathcal{O}(\alpha)$ corrections. 
\medskip

In our computation we set 
the renormalization scale, $\mu_R$, equal to the factorization scale, 
$\mu_F$, and to the  gluino mass, {\sl i.e.} 
 $\mu_R = \mu_F = m_{\tilde{g}}$.


\section{Numerical results, LHC}
\label{Sec:NumericsLHC}
For our numerical discussion we use the Standard Model parameters
quoted in Ref.~\cite{Yao:2006px}.
The value of the bottom mass 
in the $\overline{\mbox {DR}}$ scheme is computed according to
Ref.~\cite{Heinemeyer:2004xw}. We choose two different SUSY scenarios.
The first scenario is the 
SPS1a$'$ suggested by the Supersymmetry Parameters Analyses
(SPA)~\cite{AguilarSaavedra:2005pw} project.
The second  one, called SPS2, belongs to the set of Snowmass Points and
Slopes, introduced in Ref.~\cite{Allanach:2002nj}.
We obtain the parameters of the two scenarios with the help of the program
\verb|SPheno|~\cite{Porod:2003um}, 
starting  from the input parameters shown in Table~\ref{Tb:SUSYin1}. In the SPS1a$'$ 
(SPS2) scenario, the gluino mass is $608$~GeV ($784$~GeV).
\begin{table}[t]
\begin{center}
\begin{tabular}{c|c|c}
\hline
\hline
 parameter  & SPS1a$'$ & SPS2  \\
\hline
$m_{1/2}$  & $250$~GeV  &  $300$~GeV \\
$m_0$  &  $70$~GeV  & $~1450$~GeV \\
$A_0$  &  $-300$~GeV  & $0$  \\
$\mbox{sign}(\mu) $ &  $"+"$ & $"+"$  \\
$\tan \beta(M_Z)$ & $10.37$  &  $10$   \\
\hline
\hline
\end{tabular}
\caption{MSSM input parameters for the computation of the spectrum of the two scenarios considered. 
$m_{1/2}$, $m_0$ and $A_0$ are defined at the GUT scale.}
\label{Tb:SUSYin1}
\end{center}
\end{table}

\subsection{Dependence on the  SUSY scenario}
\label{SSec:results}
We compute the total hadronic cross section, 
the results are collected in Table~\ref{Tab:RES}. 
\begin{table}[t]
\begin{center}
\begin{tabular}{c|c|c|c|c}
\hline
\hline
point    & $\sigma^{\tiny\mbox{QCD, LO}}$       & $\sigma^{\tiny\mbox{QCD, LO}} + \sigma^{\tiny  \mbox{EW, NLO}}$    & $\delta$   &     $\frac{1}{\sqrt{L \cdot \sigma^{\tiny \mbox{QCD, LO}}}}$   \\
\hline
SPS1a$'$ & $6.1865(6)$ pb  & $ 6.1822(6)$ pb     &   $-0.07 \%$            & $0.13 \%$     \\\hline
SPS2    & $1.2127(1)$ pb  & $ 1.2089(1)$ pb     &   $-0.31 \%$            & $0.29 \%$     \\
\hline
\hline
\end{tabular}
\caption{Total hadronic cross section for gluino pair production at the LHC ($\sqrt{S}=14$~TeV). In the  second (third) column
  we show the $\mathcal{O}(\alpha_s^2)$  ($\mathcal{O}(\alpha^2_s + \alpha^2_s \alpha)$) contribution  
for  the points SPS1a$'$ and SPS2.  
In the fourth column the electroweak corrections relative to the LO +
NLO EW
result are given. The last column shows 
the statistical error for an integrated luminosity $L=100~\mbox{fb}^{-1}$.}
\label{Tab:RES}
\end{center}
\end{table}
The second  column shows the lowest order results. The third column shows the sum of the 
lowest order and of the $\mathcal{O}(\alpha_s^2\alpha)$
contributions. In the fourth column 
the contribution of the $\mathcal{O}(\alpha_s^2 \alpha)$ corrections 
relative to the total result is given, {\sl i.e.}  $\delta$ is defined as
\begin{displaymath}
\delta \equiv \frac{\sigma^{\tiny \mbox{EW, NLO}}}{\sigma^{\tiny \mbox{QCD, LO}}+
\sigma^{\tiny \mbox{EW, NLO}}}.
\end{displaymath}
In the last entry we give an estimate
of the statistical error based 
on an integrated luminosity  $L=
100~\mbox{fb}^{-1}$~\cite{Accomando:2005ra}.   
We do not distinguish  the results in the different renormalization schemes
since they agree within the integration error. A priori this is not guaranteed. Indeed the Rs1 scheme turns out to be unreliable in the scenarios we are considering. 
In this scheme, the finite part of the renormalization constant of the trilinear coupling, $\delta A_b^{\mbox{\tiny fin}}$, 
is comparable with the value of the trilinear coupling $A_b$ itself, {\sl i.e.} $\delta A_b^{\mbox{\tiny fin}} / A_b \sim 1$, and the 
perturbative expansion is spoiled. However, the
difference among the renormalization schemes is as small as  few
percent of the tree-level $b \bar{b}$ annihilation channel cross section.
The latter  contributions amount up to several fb, therefore  the 
variation of the results in Table~\ref{Tab:RES} is within the integration
error. 
\medskip

As one can see, in the case of the point SPS1a$'$ the electroweak corrections
are much smaller than the 
statistical uncertainty  and so 
they are not relevant. In the case of the point SPS2,  the
$\mathcal{O}(\alpha_s^2\alpha)$ corrections 
are of the same order of   
the statistical error but they are smaller than the theoretical
systematic uncertainties such 
as the uncertainty on the PDF parametrization 
($\lesssim 10 \%$)~\cite{Beenakker:1996ch}  and the factorization scale
dependence (from $3$ to $5 \%$ if $m_{\tilde{g}} \le 1$~TeV)~\cite{Kulesza:2009kq}. 
\medskip

\begin{figure}
\centering
\underline{SPS1a$'$} \\
\epsfig{file=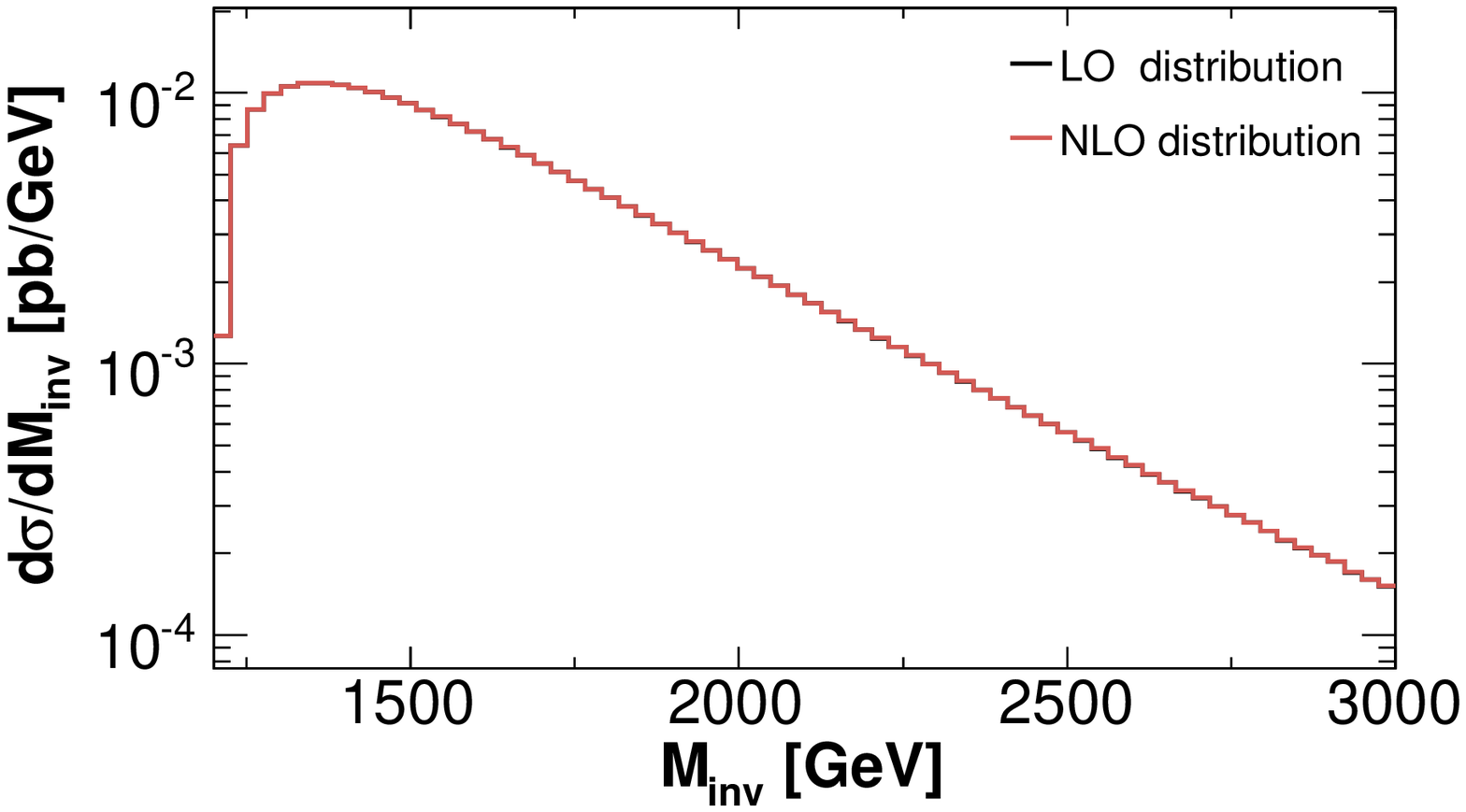, width=7cm}
\epsfig{file=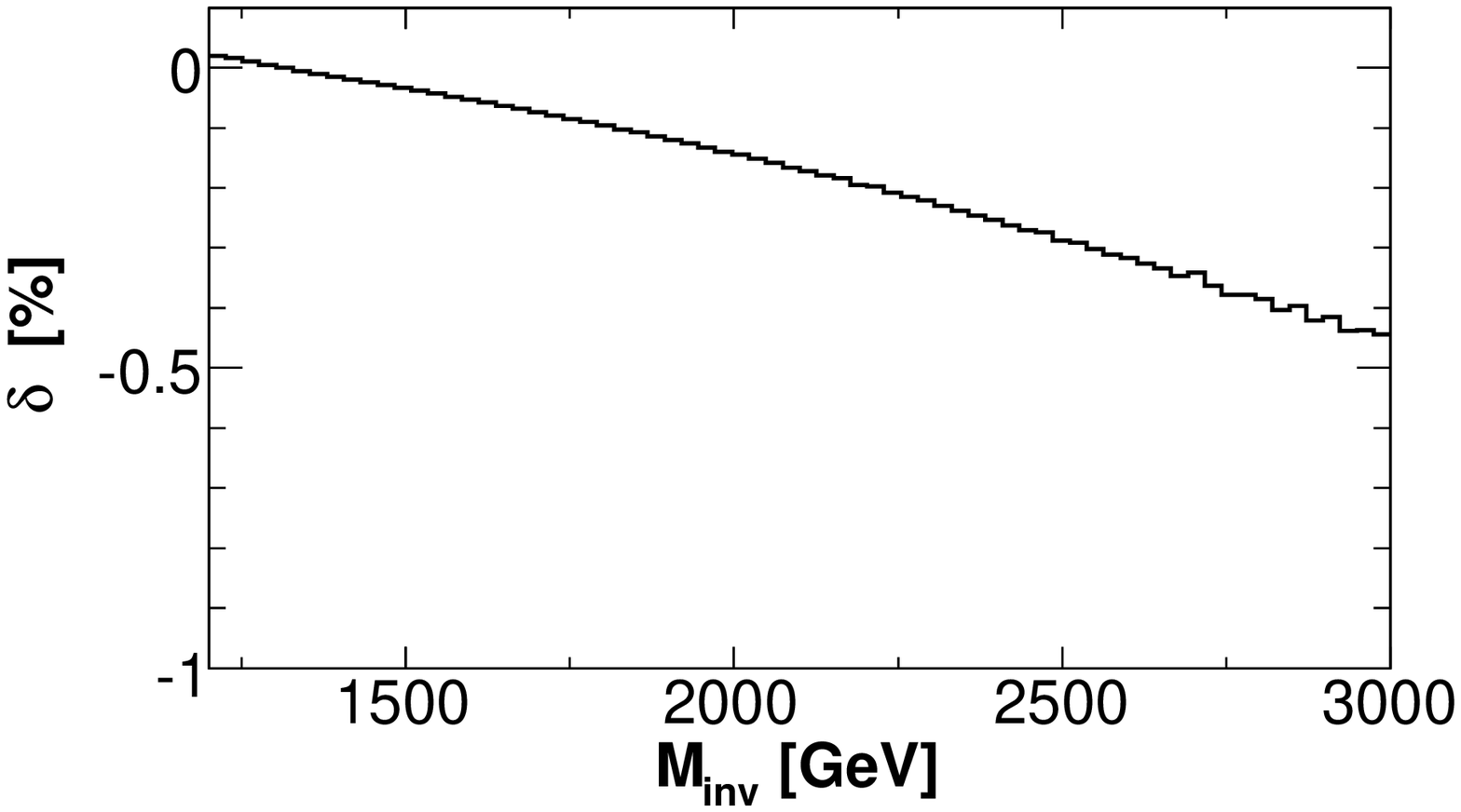, width=7cm}\\
\centering
\underline{SPS2}  \\
\epsfig{file=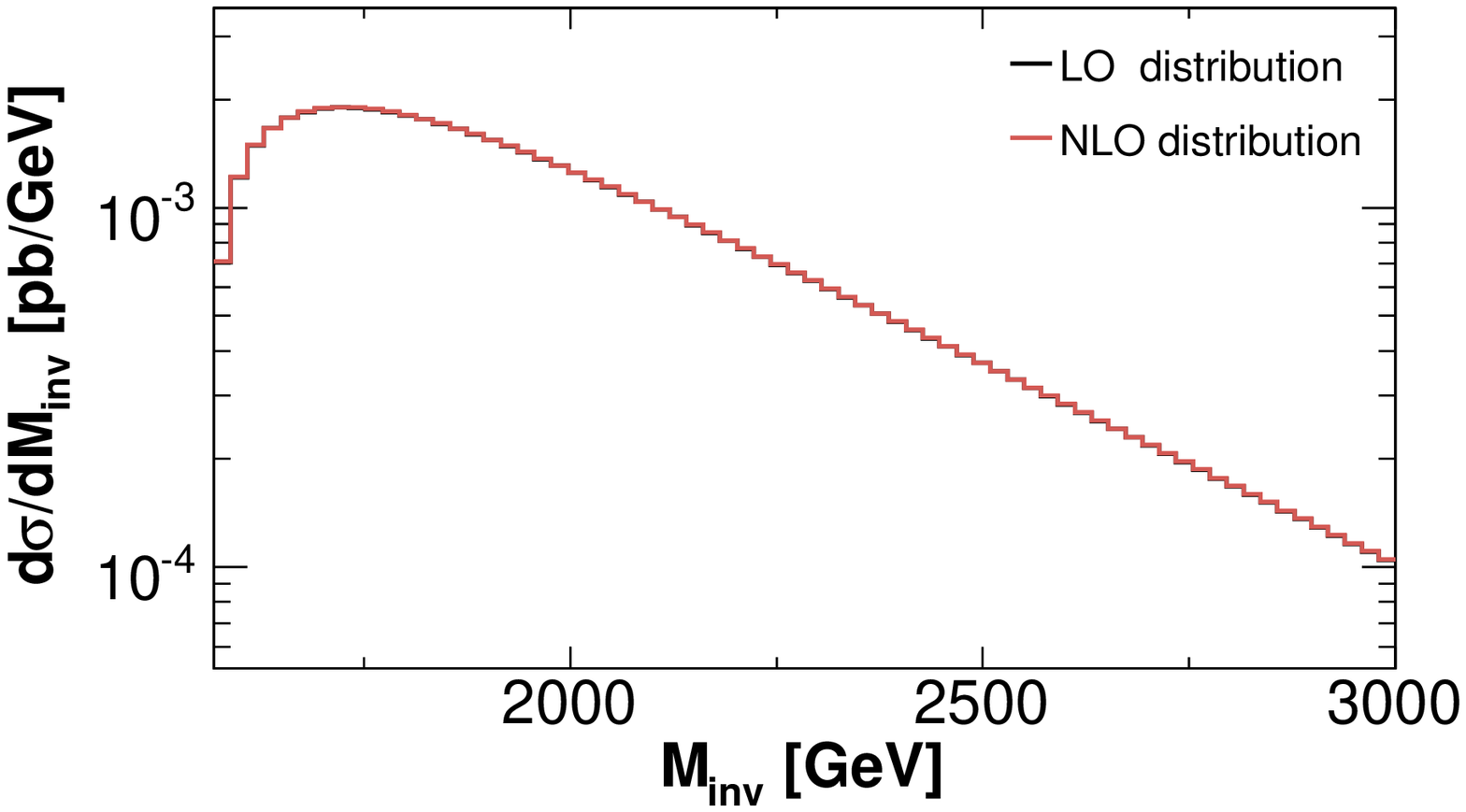, width=7cm}
\epsfig{file=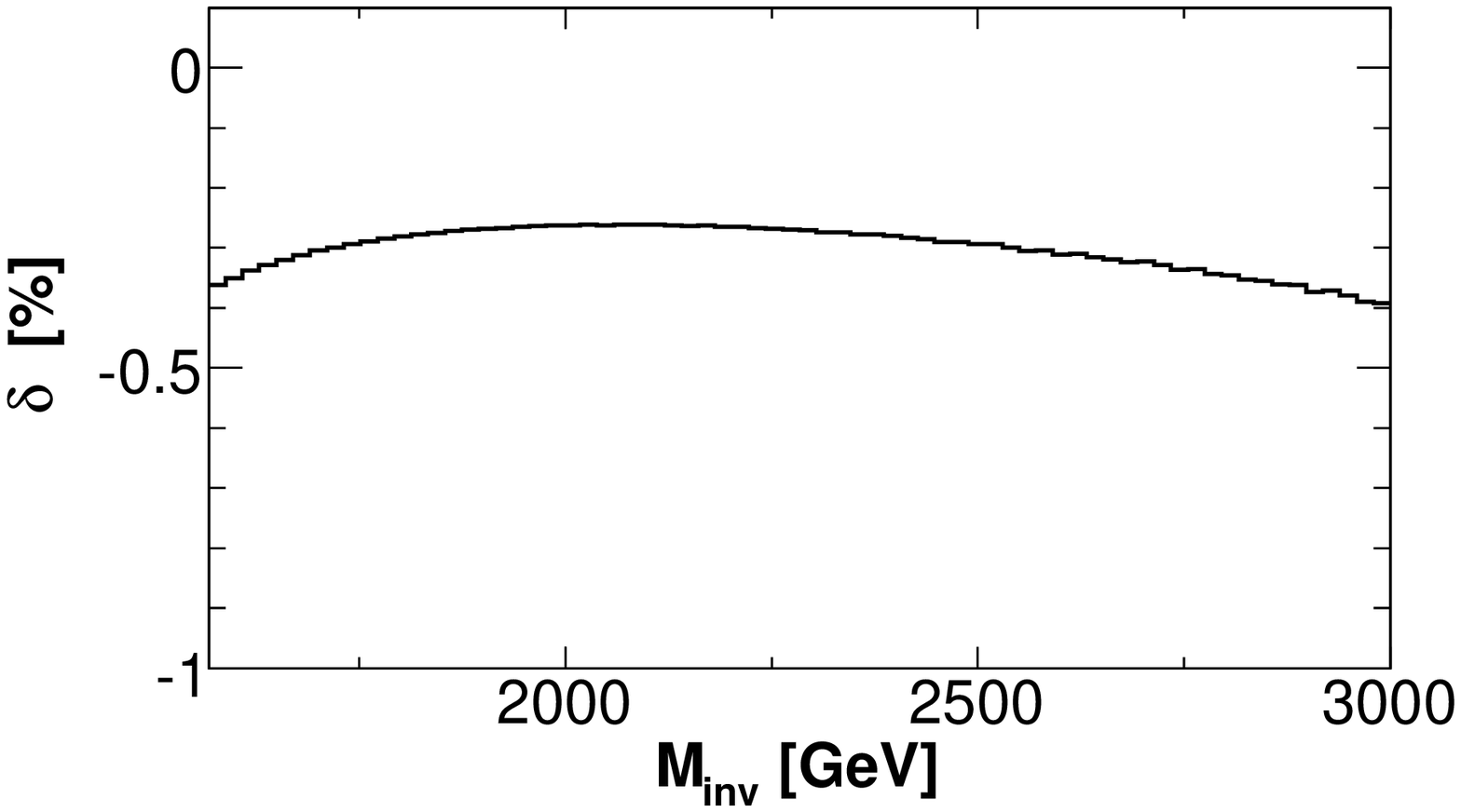, width=7cm}
\caption{Invariant mass distribution for gluino pair production at the LHC. In the left  panels 
we show the LO (black line) and  the LO + NLO EW (red line)
distribution. The two lines are indistinguishable, owing to the
smallness of the  EW contributions.  In the right panels 
the electroweak correction relative to the total result is shown.}
\label{Fig:IMa}
\end{figure}

The invariant mass distribution of the  two
gluinos is shown  in Fig.~\ref{Fig:IMa}. The 
EW corrections are small, their absolute value being at most of the order of 
$0.4\%$ of the total contribution. Moreover, these corrections do not distort
the shape of the distribution. 
\medskip

\begin{figure}
\centering
\underline{SPS1a$'$} \\
\epsfig{file=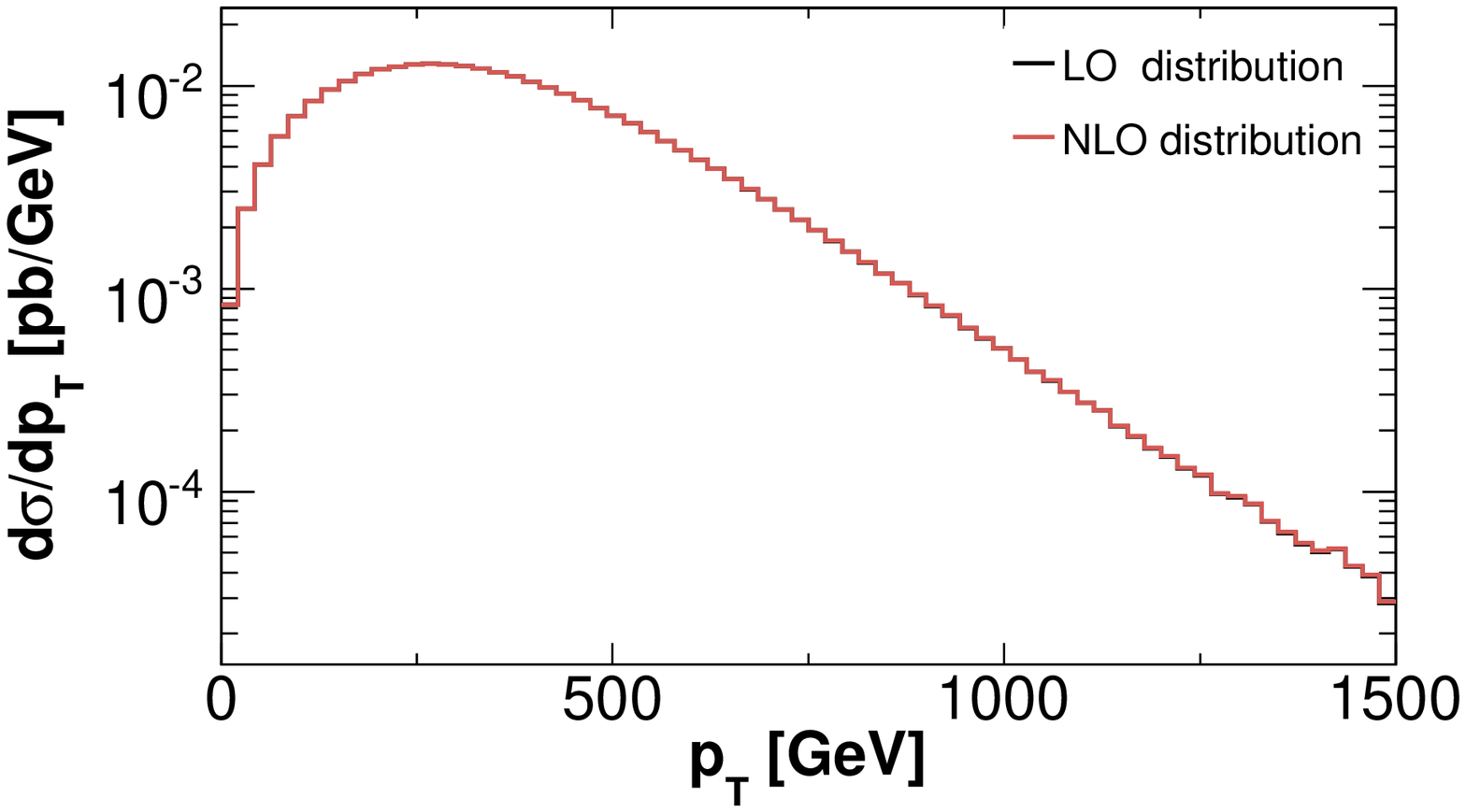, width=7cm}
\epsfig{file=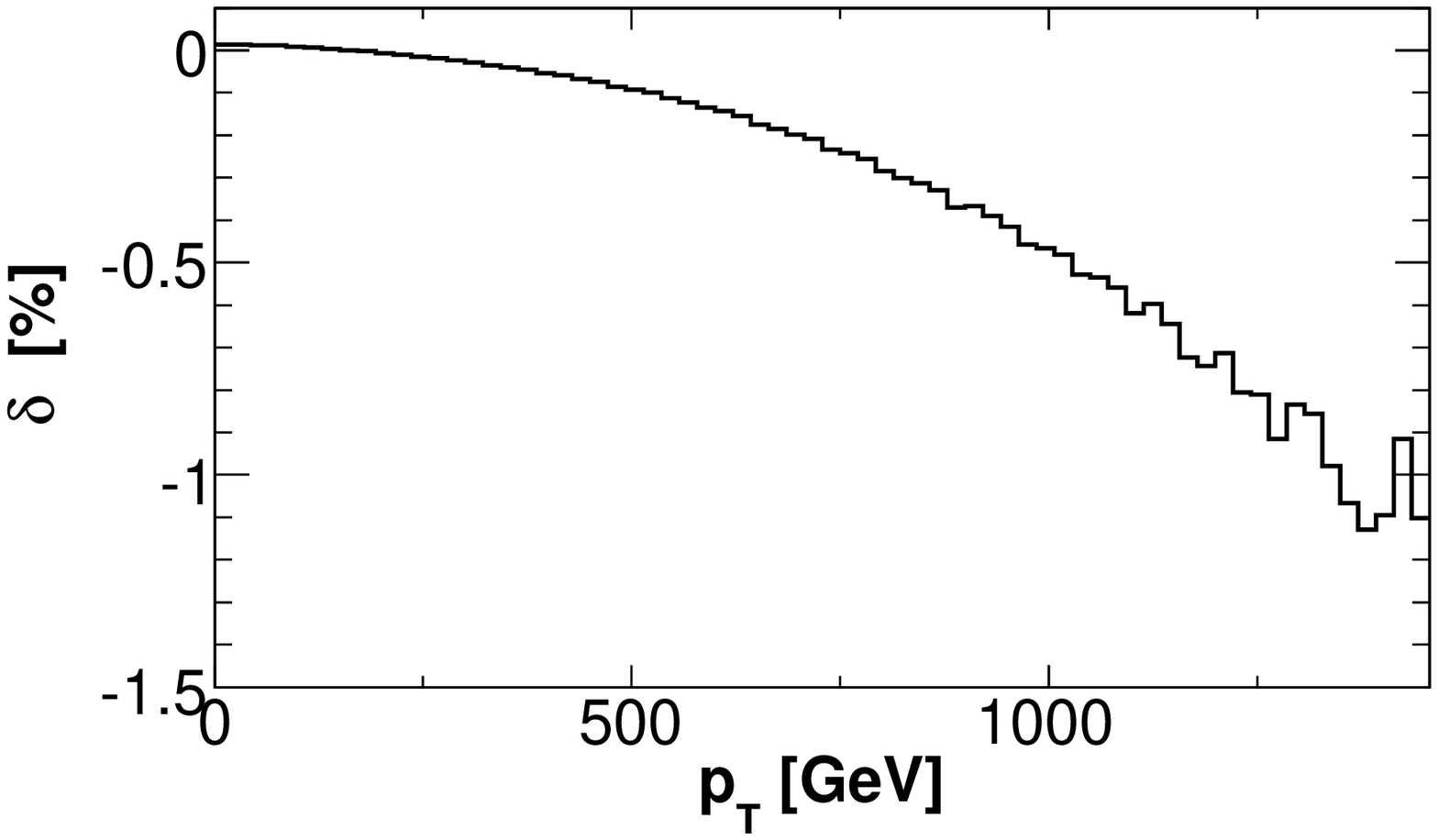, width=7cm} \\
\centering
\underline{SPS2} \\
\epsfig{file=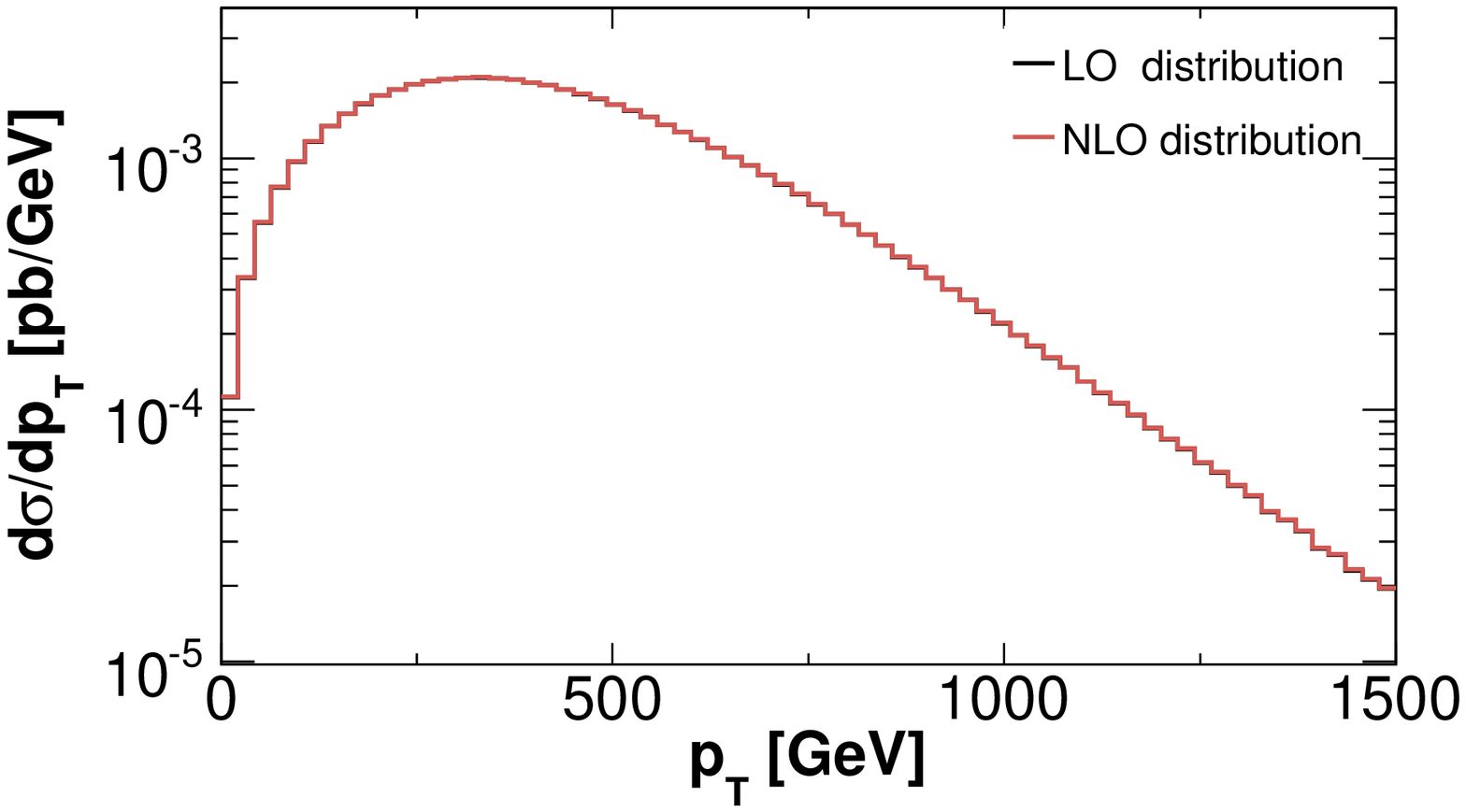, width=7cm}
\epsfig{file=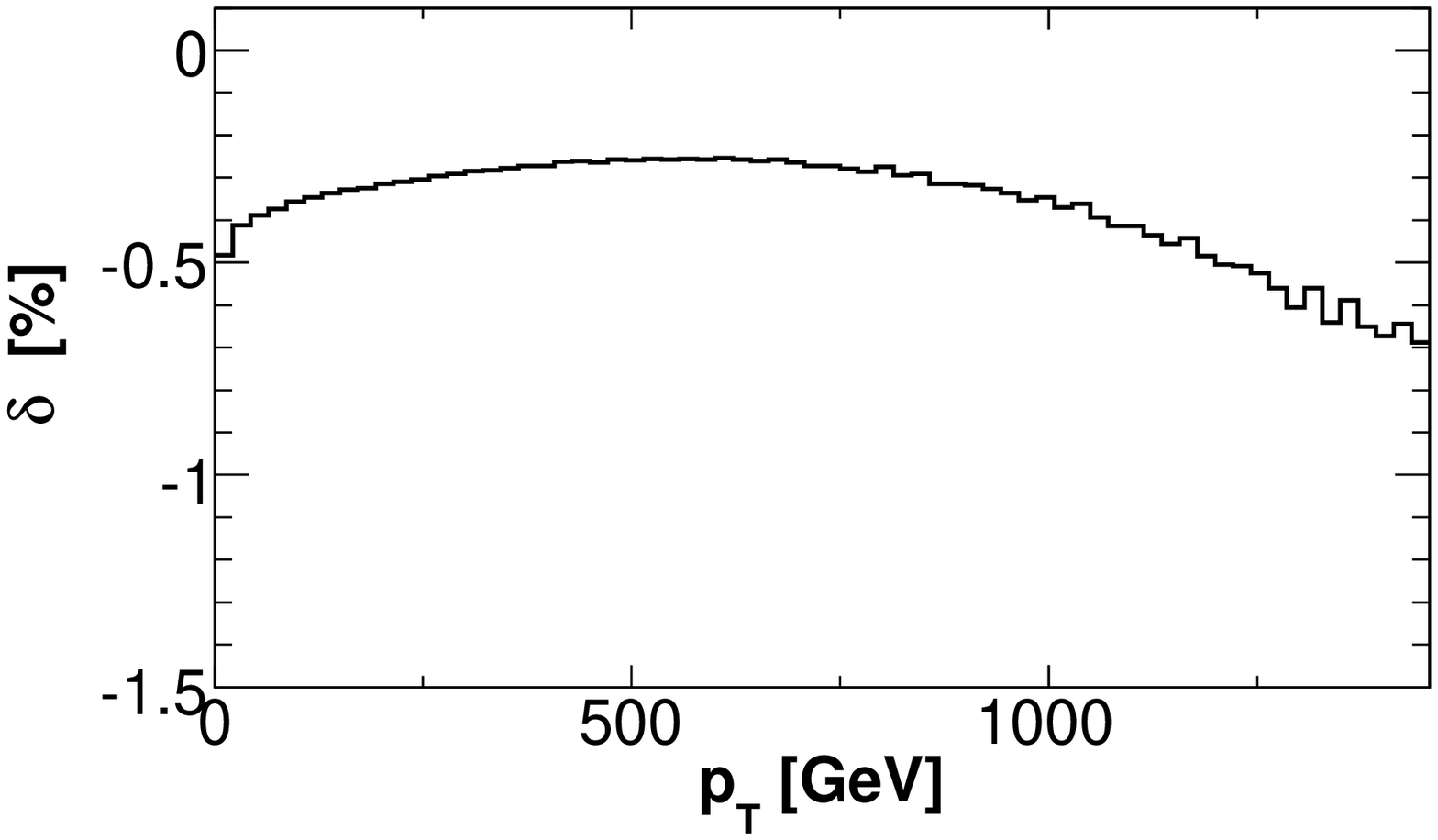, width=7cm}
\caption{Same as Fig.~\ref{Fig:IMa}, but considering the 
transverse momentum distribution.}
\label{Fig:PTa}
\end{figure}

In Fig.~\ref{Fig:PTa}  we consider the  distribution of the largest transverse momentum of 
the two gluinos, for brevity we will refer to this observable as "transverse
momentum distribution". The $\mathcal{O}(\alpha_s^2\alpha)$
corrections are rather small, the absolute value of their contribution
relative to the total result is at most $1 \%$, reaching this value  in the 
high $p_T$ region, for  $p_T \gtrsim 1500$~GeV.

\subsection{Dependence on the MSSM parameters}
In this subsection  we investigate the size of the
$\mathcal{O}(\alpha_s^2\alpha)$ corrections to gluino pair production
in a more systematic 
way, performing a scan over the parameter space of the MSSM. 
The parameters involved in the scan are the independent parameters in the second renormalization scheme. 
We suppose that all the sfermionic soft mass parameters are equal and we 
indicate their  value in the Rs2 scheme as $M_{\mbox{\tiny Susy}}$. 
The physical masses of the sfermions can be  obtained from $M_{\mbox{\tiny Susy}}$ diagonalizing the mass matrices. 
Moreover, we consider the surfaces of the parameter space characterized by $A_t = A_{\tau}$.  
With these assumptions  there are eight  independent parameters  involved in the scan, namely,
\begin{displaymath}
M_{\mbox{\tiny Susy}},~~ 
m_{\tilde{g}},~~
\mu ,~~ 
M_2,~~
A_t,~~
A_b,~~
\tan \beta ,~~
M_{A^0}. 
\end{displaymath}
The subregions of the parameter space are chosen imposing  
the exclusion limits arising 
from SUSY searches at LEP~\cite{Schael:2006cr} and at the
Tevatron~\cite{Abazov:2006bj}, and the bound on the mass of the
light Higgs boson. The physical mass of the light Higgs boson  has been computed 
using \verb|FeynHiggs 2.5.1|~\cite{Heinemeyer:1998yj,Frank:2006yh,Degrassi:2002fi}. 
Moreover, each point in the selected regions  fulfills the condition
$|\Delta \rho | \leq 0.025$, $\Delta \rho$ being the dominant SUSY
corrections to the electroweak $\rho$ parameter, corrections arising
from  top and bottom squarks contributions.
\medskip

We perform four different scans. In each scan we select two parameters and we
study the dependence 
of the quantity $\Delta$, 
\begin{displaymath}
\Delta\equiv \frac{\sigma^{\mbox{\tiny EW, NLO}}}{\sigma^{\mbox{\tiny QCD, LO}}+\sigma^{\mbox{\tiny EW, NLO}}} \cdot 100.
\end{displaymath}
We repeat each of these scans for different values of another
pair of parameters, while the remaining four are fixed to their SPS1a$'$
values. Here there is a brief discussion on the results of these scans.
%
\subsubsection*{Scan over \boldmath{$A_t$}  and \boldmath{$A_b$}}
\begin{figure}
\vspace{-2cm}
\centering
\epsfig{file=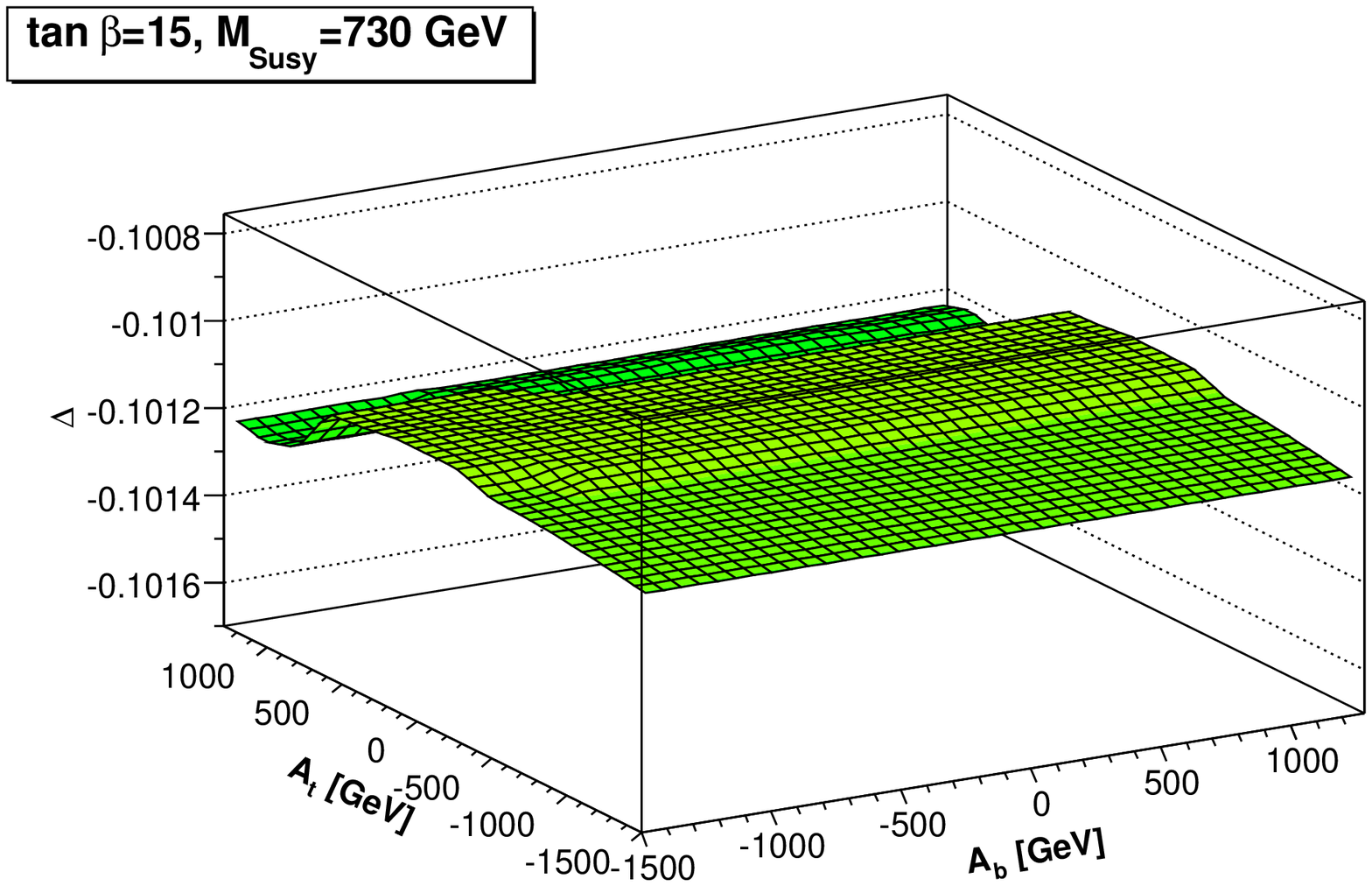, width=7cm}
\epsfig{file=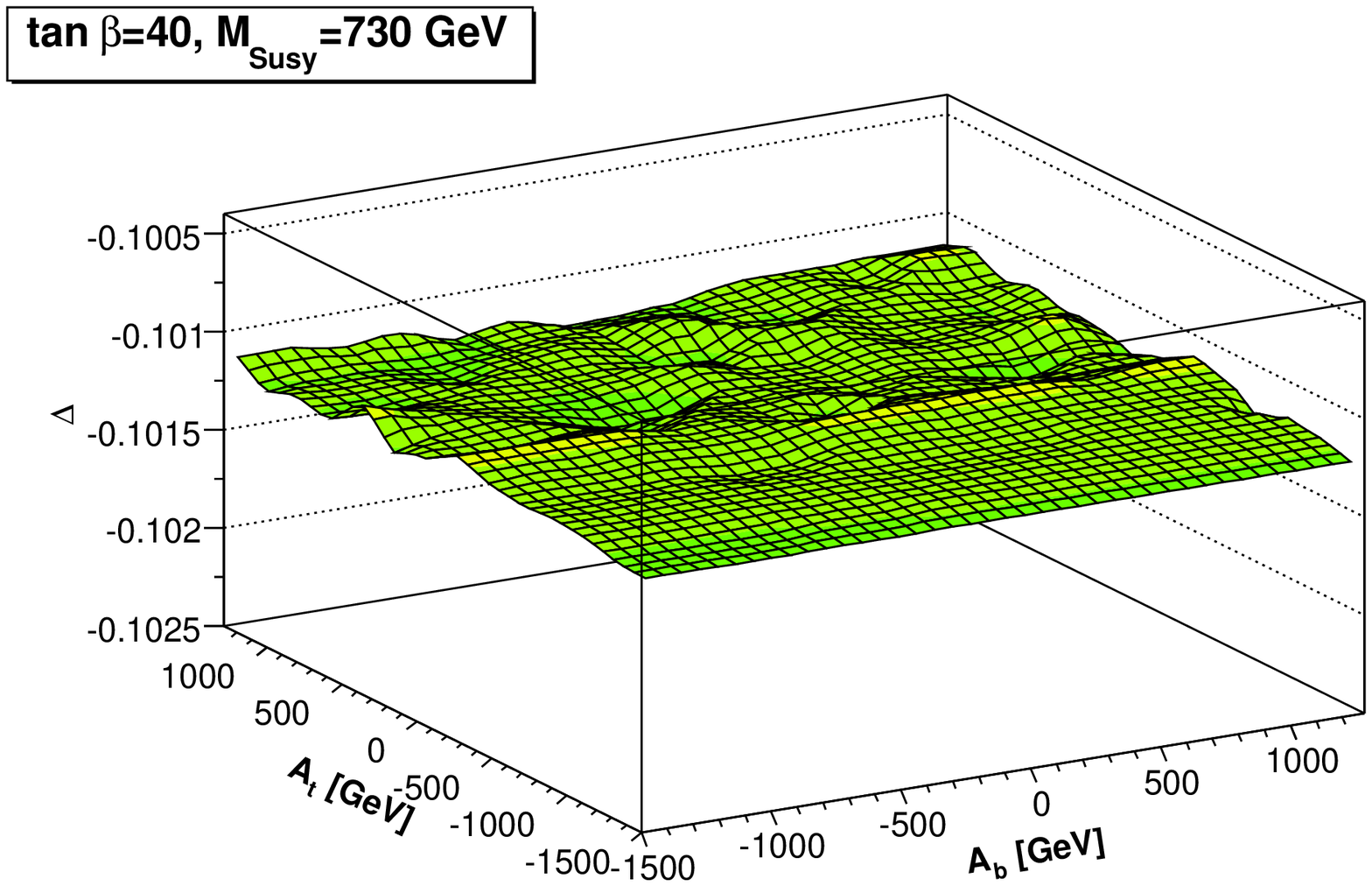, width=7cm}
\caption{$ \Delta= \, \sigma^{\mbox{\tiny EW, NLO}} / (\, \sigma^{\mbox{\tiny QCD, LO}} + \sigma^{\mbox{\tiny EW, NLO}} \, )\cdot 100$  as a function of $A_t$  and $A_b$ for different values 
of $\tan \beta$ and $M_{\mbox{\tiny Susy}}$. The other parameters are fixed to their 
SPS1a$'$ values.}
\label{Fig:Scan1}
\vspace{2.2cm}
\centering
\epsfig{file=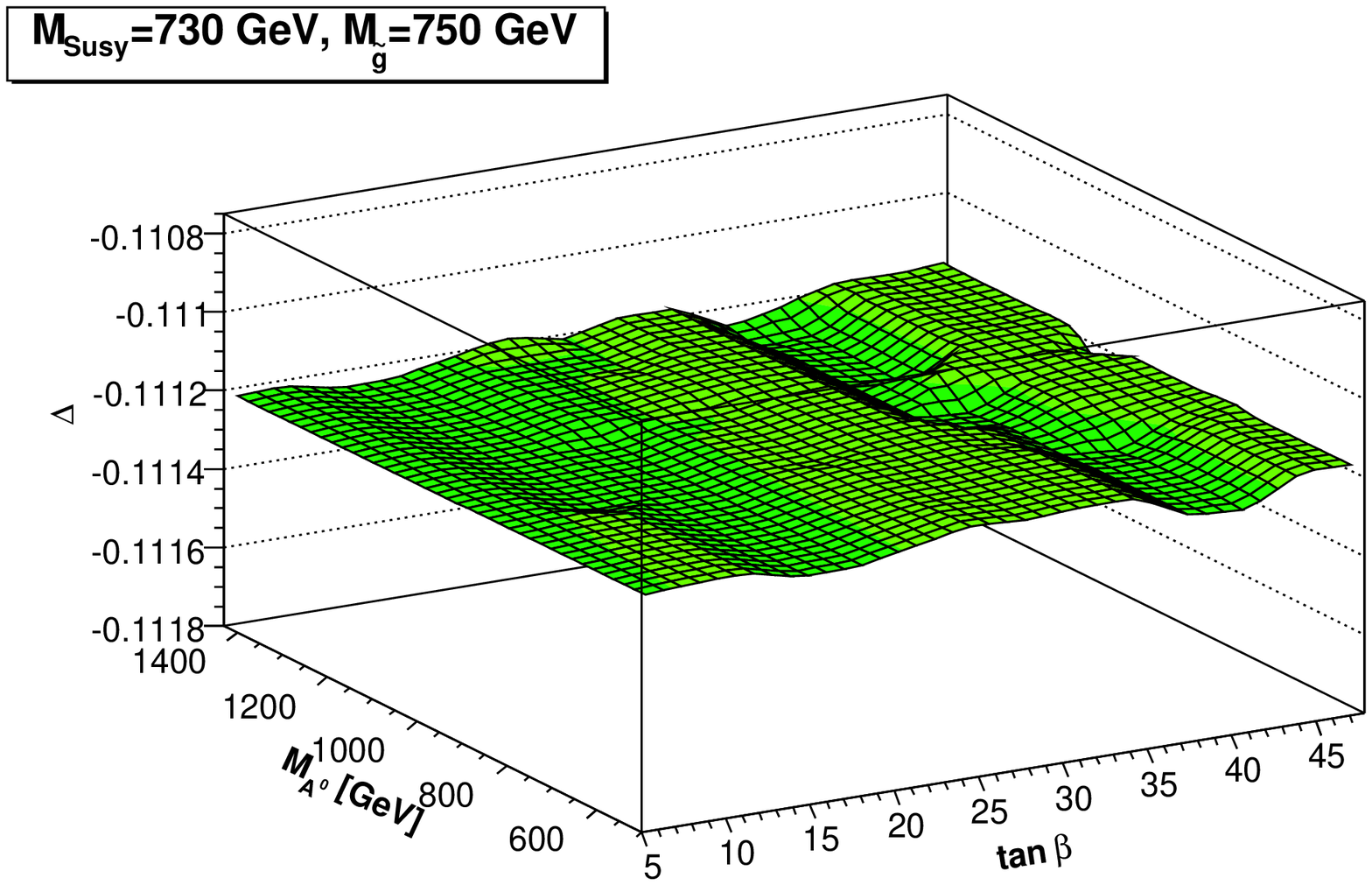, width=7cm}
\epsfig{file=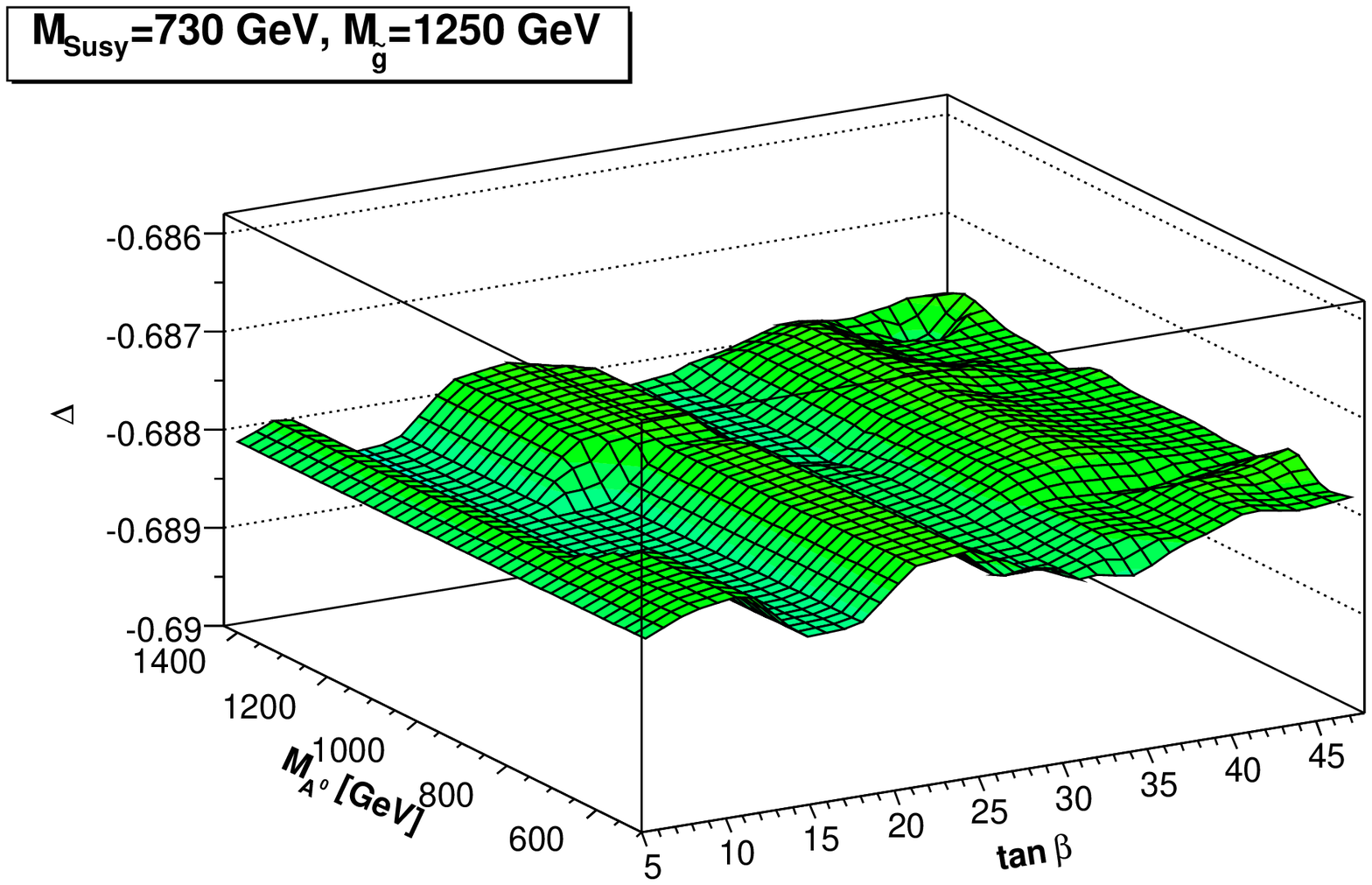, width=7cm}
\caption{$ \Delta= \, \sigma^{\mbox{\tiny EW, NLO}} / (\, \sigma^{\mbox{\tiny QCD, LO}} + \sigma^{\mbox{\tiny EW, NLO}} \, ) \cdot 100$    as a function of $\tan \beta$  and $M_{A^0}$ for 
different values of $m_{\tilde{g}}$ and $M_{\mbox{\tiny Susy}}$. The other parameters 
are fixed to their SPS1a$'$ values.}
\label{Fig:Scan3}
\end{figure}

The results of this scan are displayed in Fig.~\ref{Fig:Scan1}. As expected,
$\Delta$ is quite independent 
on the parameter $A_t$ which enters 
in the virtual correction of the process $b \overline{b} \to
\tilde{g}\tilde{g}$ and in the definition of the mass of the top squarks. 
This feature  is more evident for large $\tan \beta$ values.  $\Delta$
varies only by an amount of the 
order of few percent for a variation of  
$A_b$ and $A_t$  over a quite broad range
(from $-1500$ to $1500$ GeV ). Note that in the whole subregion considered the
absolute value of $\Delta$ is of $\mathcal{O}(0.1)$.
\subsubsection*{Scan over \boldmath{$\tan \beta$}  and \boldmath{$M_{A^0}$} }
As can be inferred from Fig.~\ref{Fig:Scan3}, the dependence of $\Delta$ on
$(\tan \beta, M_{A^0})$ strongly varies 
for different values of $m_{\tilde{g}}$ and 
$M_{\mbox{\tiny Susy}}$. 
As a general result the overall dependence is mild for each value
of $(m_{\tilde{g}}, M_{\mbox{\tiny Susy}})$. In 
all cases 
the value of $|\Delta|$ is at most of  the order of $0.7$.
\subsubsection*{Scan over \boldmath{$\mu$} and \boldmath{$M_2$}}
\begin{figure}
\vspace{1cm}
\centering
\epsfig{file=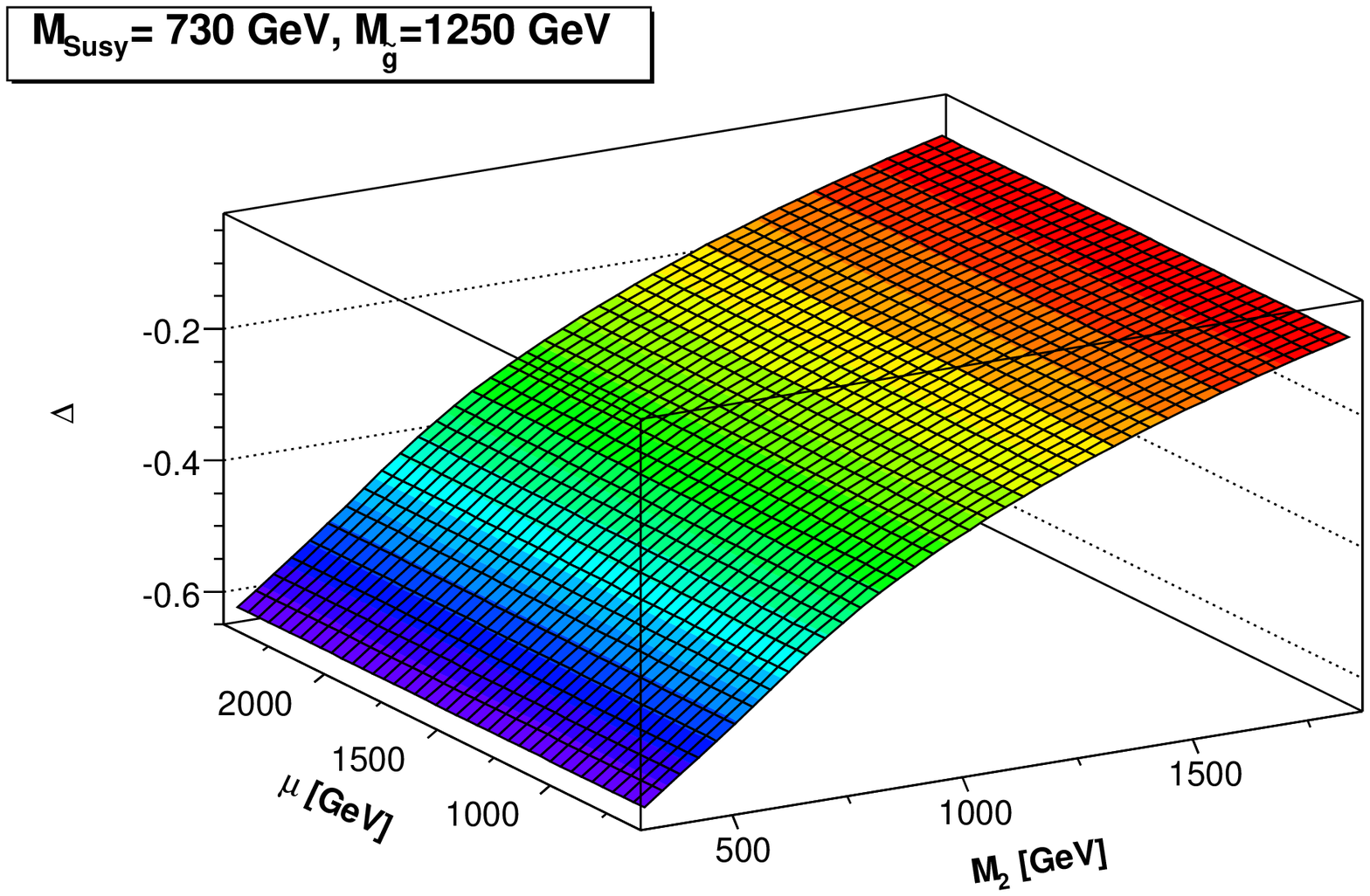,  width=7cm}
\epsfig{file=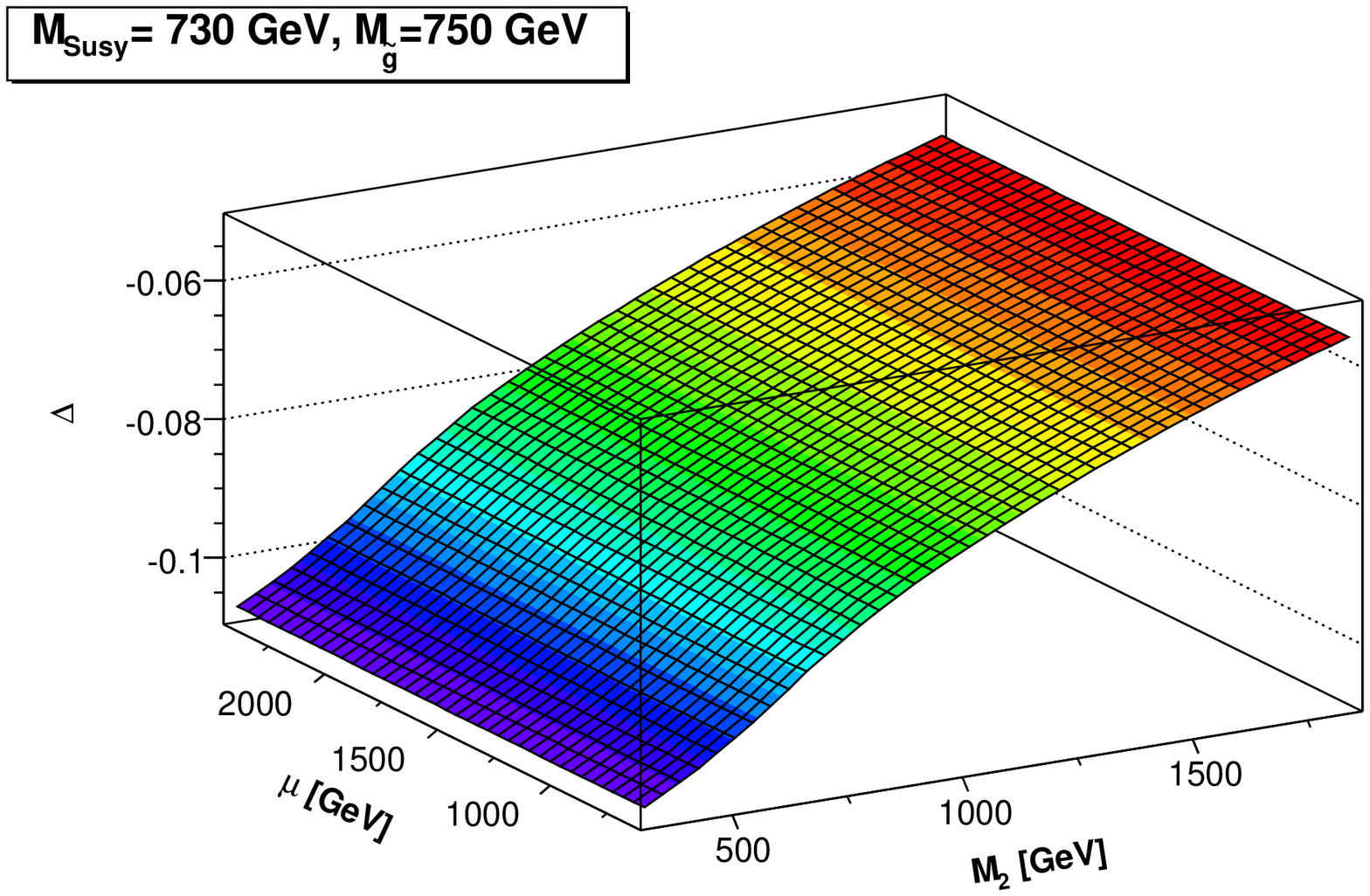, width=7cm}
\vskip 1cm
\centering
\epsfig{file=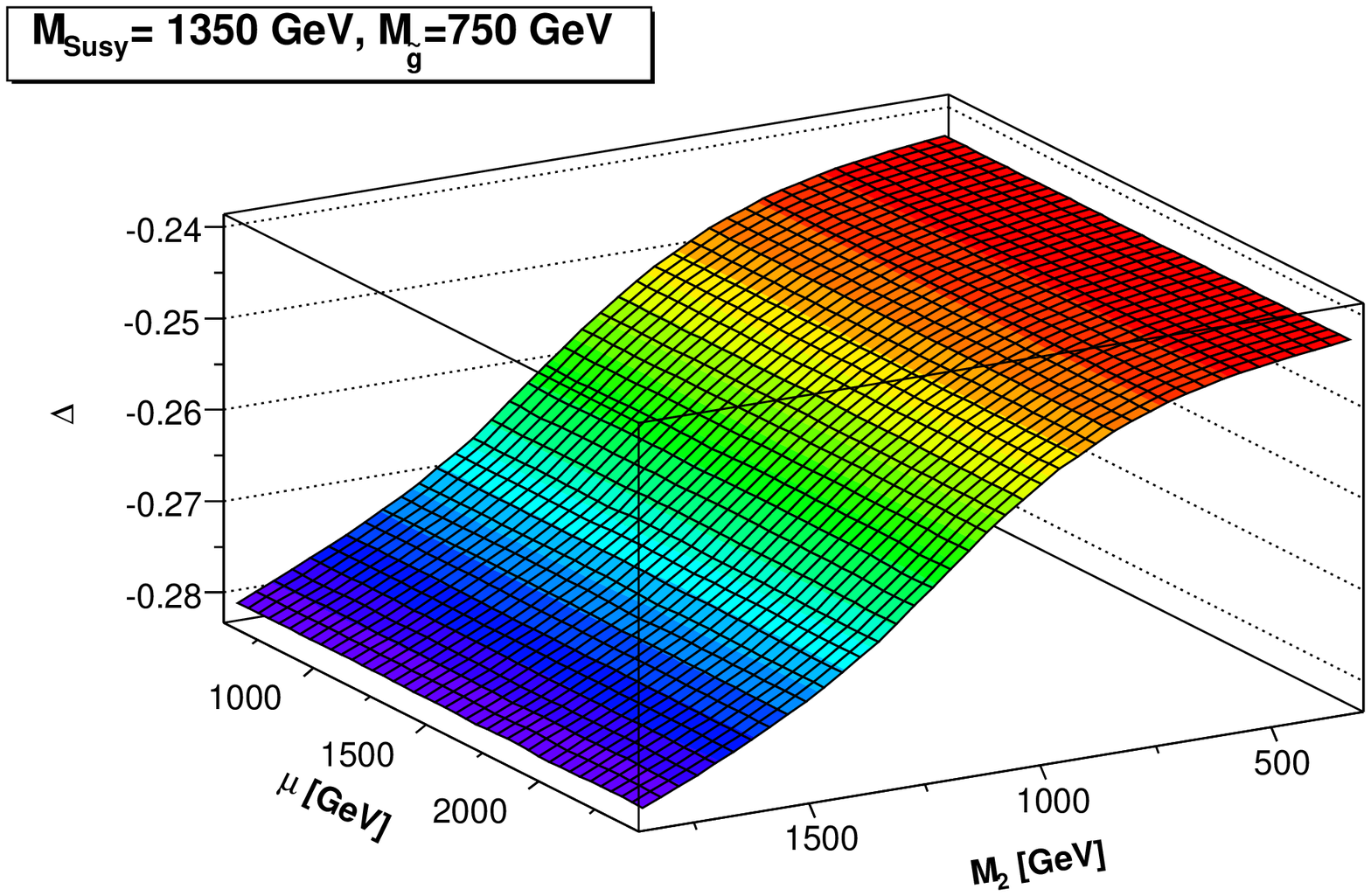,  width=7cm}
\epsfig{file=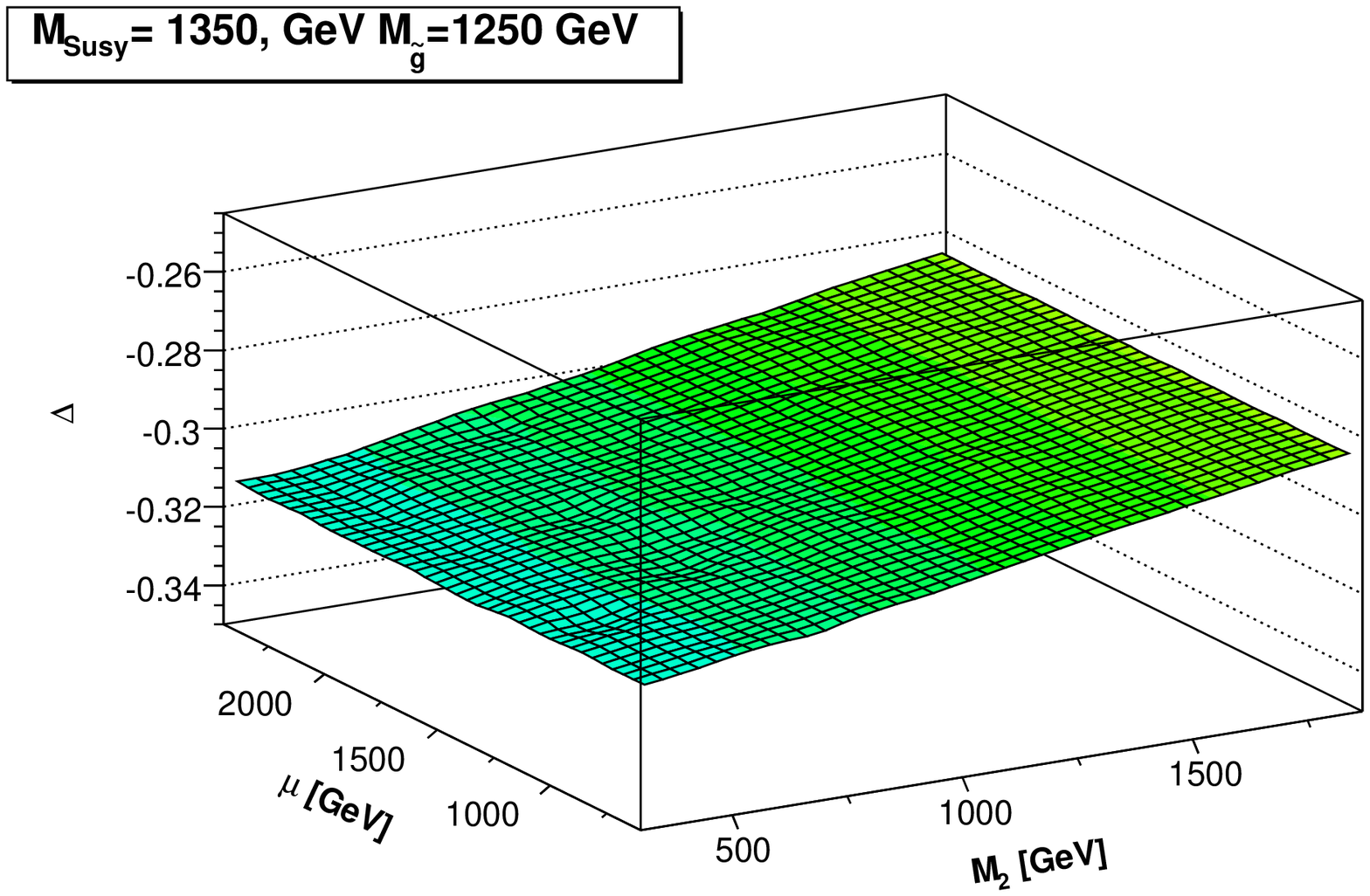, width=7cm}
\caption{$ \Delta= \, \sigma^{\mbox{\tiny EW, NLO}} / (\, \sigma^{\mbox{\tiny QCD, LO}} + \sigma^{\mbox{\tiny EW, NLO}} \, ) \cdot 100$  as a 
function of $\mu$  and $M_2$ for different values 
of $m_{\tilde{g}}$ and $M_{\mbox{\tiny Susy}}$. The other parameters 
are fixed to their SPS1a$'$ values.}
\label{Fig:Scan2}
\end{figure}

As displayed in Fig.~\ref{Fig:Scan2}, $\Delta$ is almost independent on
$\mu$ for each value of the 
pair $(m_{\tilde{g}}, M_{\mbox{\tiny Susy}})$ while the 
dependence on $M_2$ is more important and particularly pronounced when $m_{\tilde{g}} =1250$~GeV and  $M_{\mbox{\tiny Susy}}=730$~GeV. 
In the case of the last three plots the value of $\Delta$ is of order $-0.3$
to $-0.1$, while  in the 
first plot, characterized by   
$m_{\tilde{g}}\sim 2 \cdot M_{\mbox{\tiny Susy}}$, the value of $\Delta$ is enhanced for
small values of $M_2$ 
reaching  the value of $-0.65$. \\
\noindent
Notice that the mass of the lightest neutralino and chargino is almost
independent on the value of $\mu$ but 
varies strongly as $M_2$ varies, growing as the value 
of this parameter grows. So this enhancement occurs when  charginos$/$neutralinos
are much lighter than the gluino.
\subsubsection*{Scan over \boldmath{$M_{\mbox{\tiny Susy}}$} and \boldmath{$m_{\tilde{g}}$}}
\begin{figure}
\vspace{1cm}
\centering
\epsfig{file=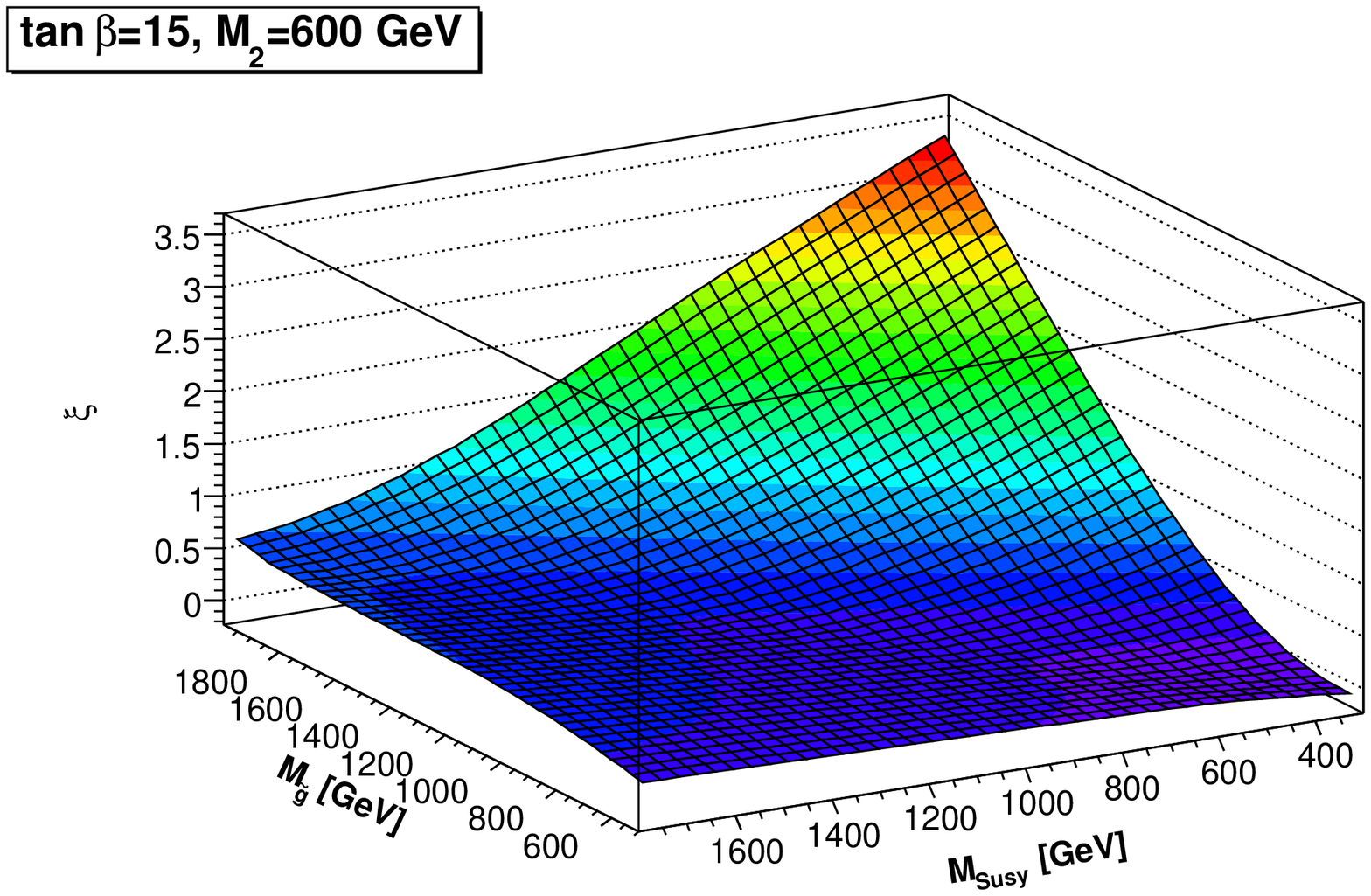, width=7cm}
\epsfig{file=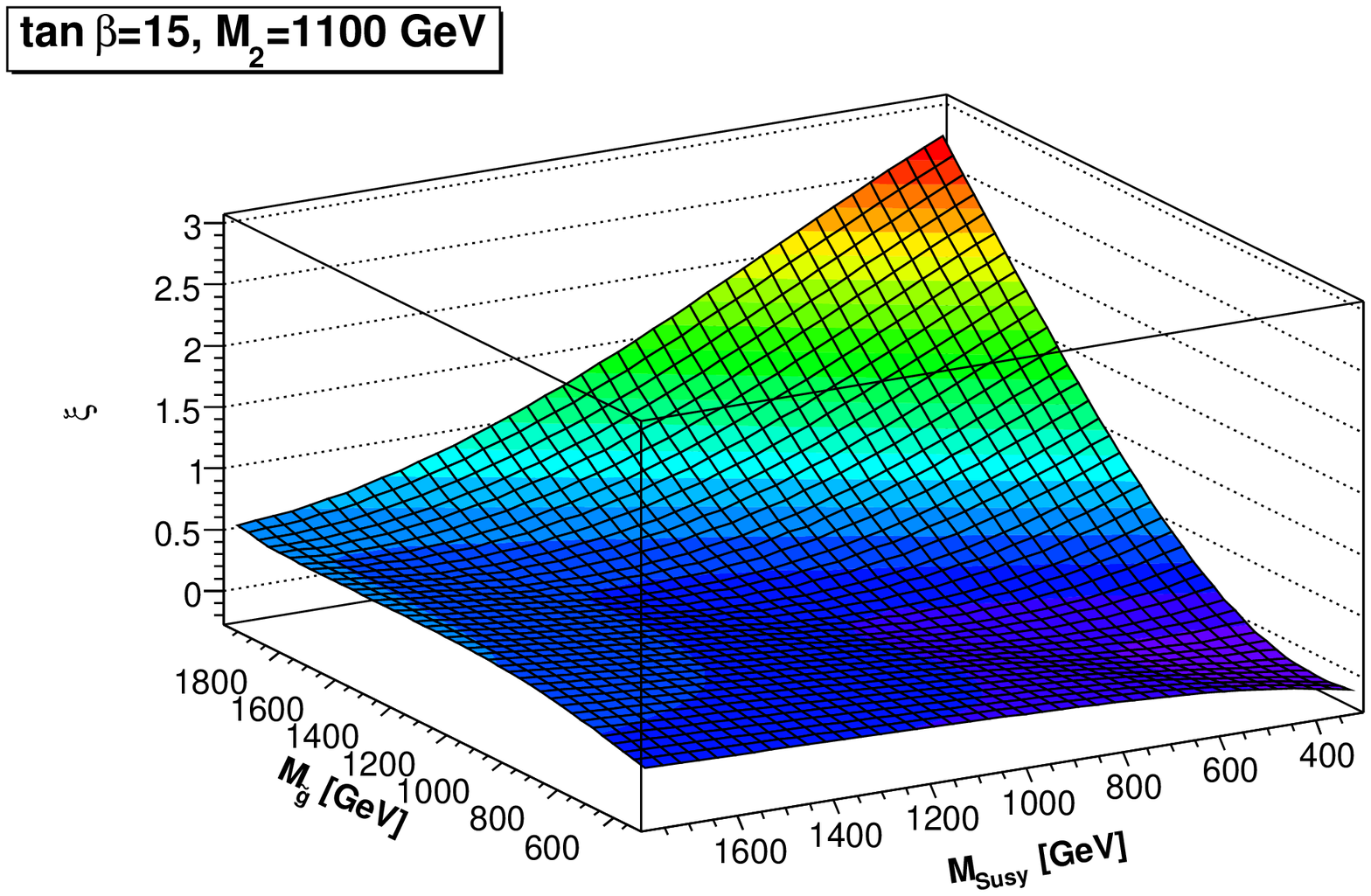, width=7cm}
\caption{$\xi \equiv -  \Delta= \, -\sigma^{\mbox{\tiny EW, NLO}} / (\, \sigma^{\mbox{\tiny QCD, LO}} + \sigma^{\mbox{\tiny EW, NLO}} \, ) \cdot 100$ 
 as a function of
  $M_{\mbox{\tiny Susy}}$  and $m_{\tilde{g}}$ for different 
values of $M_2$ and $\tan \beta$. The other parameters 
are fixed to their SPS1a$'$ values.}
\label{Fig:Scan5}
\end{figure}

In this scan we investigate the dependence of $\Delta$ on $m_{\tilde{g}}$ and
$M_{\mbox{\tiny Susy}}$, which is expected to 
be the most important because
of the dependence of the lowest order cross section on these parameters. We consider the variation of $\Delta$ 
as a function of $(m_{\tilde{g}}, M_{\mbox{\tiny Susy}})$ 
for different values of $M_2$ and $\tan \beta$, see 
Fig.~\ref{Fig:Scan5}. Note that we plot $\xi \equiv - \Delta$ instead of $\Delta$.
As a general feature 
$\xi$  increases as $m_{\tilde{g}}$ increases and as $M_{\mbox{\tiny Susy}}$ decreases. The behaviour of $\xi$
as a function of $M_{\mbox{\tiny Susy}}$ and $m_{\tilde{g}}$ is affected by
the value of $M_2$ 
being enhanced for smaller 
values of this parameter. In particular $\xi \sim 3$ in the region
$m_{\tilde{g}} \ge  1600$~GeV, $M_{\mbox{\tiny Susy}} \le 500$~GeV. \\
The  enhancement of the EW corrections is related to the increasing importance of the
$q\bar{q}$ annihilation channel when the production threshold becomes
higher. Indeed, the minimal value of the parton's momentum fraction rises as the gluino mass rises.  Since
the relative importance of the (anti-)quark PDF
increases as the momentum fraction of the (anti-)quark increases, the EW
corrections grow as the mass of the gluino grows. The relative importance of the EW contributions is more pronounced when $M_{\mbox{\tiny Susy}}$ is
small owing to the presence of tree-level diagrams with
squarks exchanged in the $t$ and $u$ channel, {\sl c.f.} Fig.~\ref{Fig:TREE}, which are enhanced when the squark masses decrease.

\section{Numerical results, Tevatron}
\label{Sec:NumericsTEV}
The EW contributions to gluino pair production are expected to be more important at the Tevatron than at the LHC, owing to the 
enhancement of the quark--anti-quark annihilation channels with respect to the gluon fusion channel. Therefore, it is worth to estimate 
the impact of the EW contributions to gluino pair production at the Tevatron, {\sl i.e.} to the process
\begin{equation}
\label{Eq:ProcessGLUTev}
P~\overline{P} \to \tilde{g}~\tilde{g}~X.
\end{equation}
The previous analysis can be easily extended to~(\ref{Eq:ProcessGLUTev}), provided that the definition of the 
luminosity, Eq.~(\ref{Eq:Lumi}), is replaced by 
\begin{equation}
\frac{dL_{ij}}{d\tau}(\tau)= \frac{1}{1+\delta_{ij}}~
\int_{\tau}^1 \frac{dx}{x} \left [ f_{i|P}(x) f_{j|\overline{P}}\left(\frac{\tau}{x}\right)+ 
f_{j|P}\left(\frac{\tau}{x}\right)f_{i|\overline{P}}(x) \right ].
\end{equation}

For numerical evaluation, 
we focus on two different points of the MSSM parameter space,
referred to  as TP1 and TP2 respectively. These points belong
to the region of the parameter space of the MSSM used in 
the data analysis made by CDF and D0
collaborations~\cite{Abbott:1999xc,Affolder:2001tc,:2007ww}. 
We obtain the parameters in these scenarios with the help of \verb|SPheno|, starting from the input 
parameters at the GUT scale described in Table~\ref{Tb:SUSYinT}.
\begin{table}[t]
\begin{center}
\begin{tabular}{c|c|c}
\hline
\hline
 parameter          & TP1         & TP2  \\
\hline
$m_{1/2}$           & $200$~GeV   &  $500$~GeV \\
$m_0$               &  $130$~GeV  & $120$~GeV \\
$A_0$               &  $0$        & $0$  \\
$\mbox{sign}(\mu) $ &  $"-"$      & $"-"$  \\
$\tan \beta(M_Z)$       & $3$         &  $3$   \\
\hline
\hline
\end{tabular}
\caption{MSSM input parameters for the computation of the spectrum of
  the scenarios TP1 and TP2.}
\label{Tb:SUSYinT}
\end{center}
\end{table}
These points are compatible with the experimental limits set by the analysis made 
by the D0 collaboration~\cite{:2007ww}. In particular, the first one corresponds to a scenario in which 
the gluino is heavier than the squarks ($m_{\tilde{g}} \sim 500 $~GeV,
$m_{\tilde{q}\ne \tilde{t}}\sim 460$~GeV), 
while the second one
describes a scenario characterized by  a light gluino
($m_{\tilde{g}}\sim 340$~GeV, 
and  $m_{\tilde{q} \ne \tilde{t}}\sim 550$~GeV).
\medskip

In Table~\ref{Tab:REST} we show the total hadronic cross $\mbox{section}$ in the
two points considered.
\begin{table}[t]
\begin{center}
\begin{tabular}{c|c|c|c|c}
\hline
\hline
point    & $\sigma^{\tiny \mbox{QCD, LO}}$                 & $\sigma^{\tiny \mbox{QCD, LO}} + \sigma^{\tiny \mbox{EW, NLO}}$    
&  $\delta$ 
&   $\frac{1}{\sqrt{L \cdot \sigma^{\tiny \mbox{QCD, LO}}}}$     \\
\hline
TP1 & $0.16714(1) $ fb         & $0.16691(1) $ fb   &  $-0.14 \%$     &  $61~\%$                                                 \\
TP2 & $0.048864(3)$ pb         & $0.048256(4)$ pb   &  $-1.26 \%$   &    $3.6~\%$                                                \\
\hline
\hline
\end{tabular}
\caption{Same as Table~\ref{Tab:RES}, 
  but considering gluino pair production
  at the Tevatron, {\sl i.e.} the process $P \overline{P} \to \tilde{g} \tilde{g} X$ at $\sqrt{S}=1.96$~TeV, and different SUSY scenarios.
In this case,  the last column shows the statistical error for an integrated luminosity    $L = 2 \times 8~\mbox{fb}^{-1}$.}
\label{Tab:REST}
\end{center}
\end{table}
We use the same notation as in Table~\ref{Tab:RES}. In the case of the point TP1, the size of the electroweak
corrections is so small that they will  
not be visible at the expected final integrated luminosity 
$L = 2 \times 8~\mbox{fb}^{-1}$ . In the case of the point TP2 we obtain a relative statistical error of order $4 \%$   
which is  three times bigger than the size of the electroweak contributions. Moreover it is worth to notice that
the systematic uncertainties affecting SUSY searches at the Tevatron 
are typically greater than $1 \%$. For instance, Ref.~\cite{:2007ww} claims that   
the $\mu_F$ dependence of the total cross section gives an error from $15$ to $20 \%$.
\medskip

\begin{figure}
\centering
\underline{TP1}\\
\epsfig{file=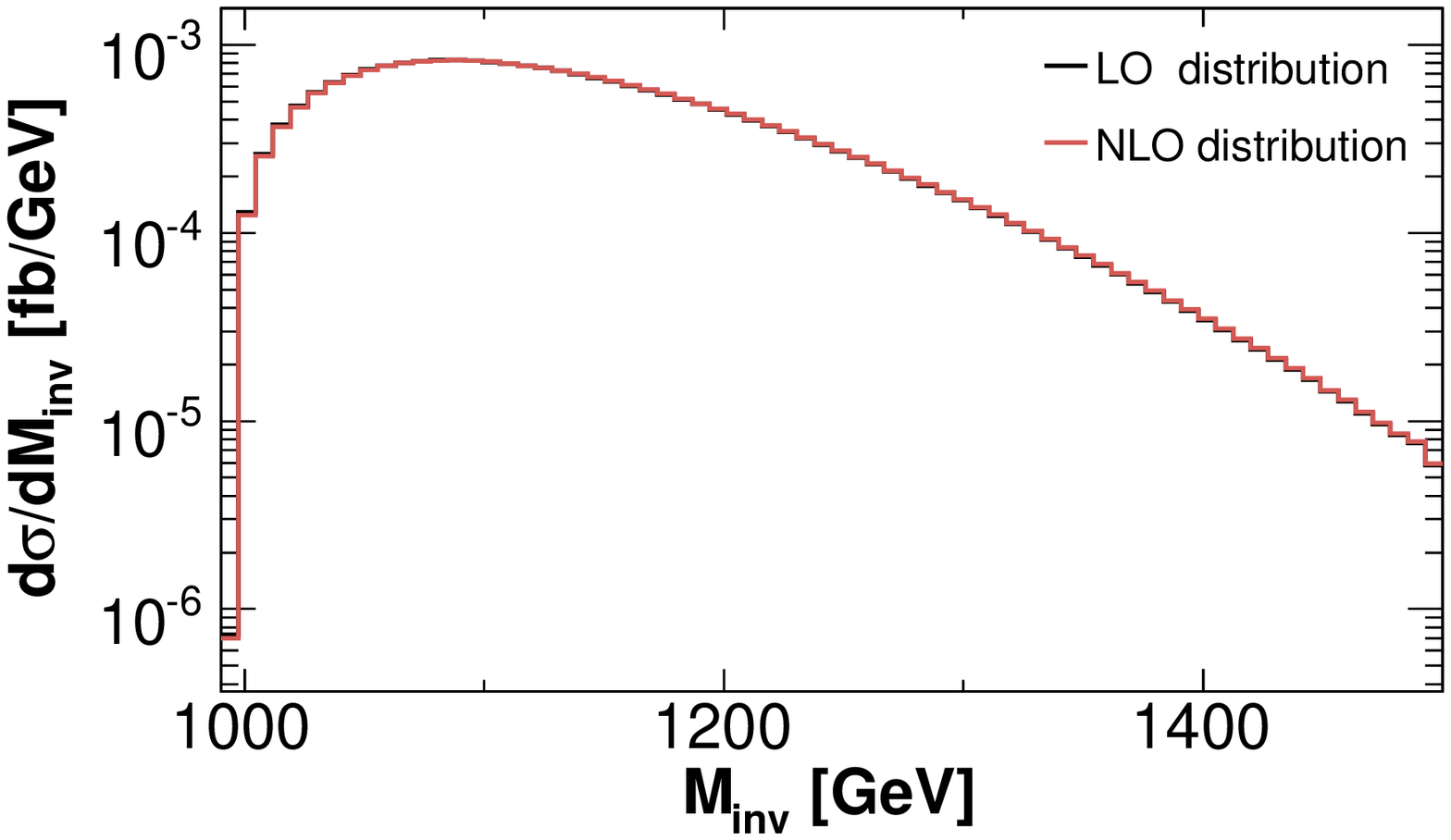, width=7cm}
\epsfig{file=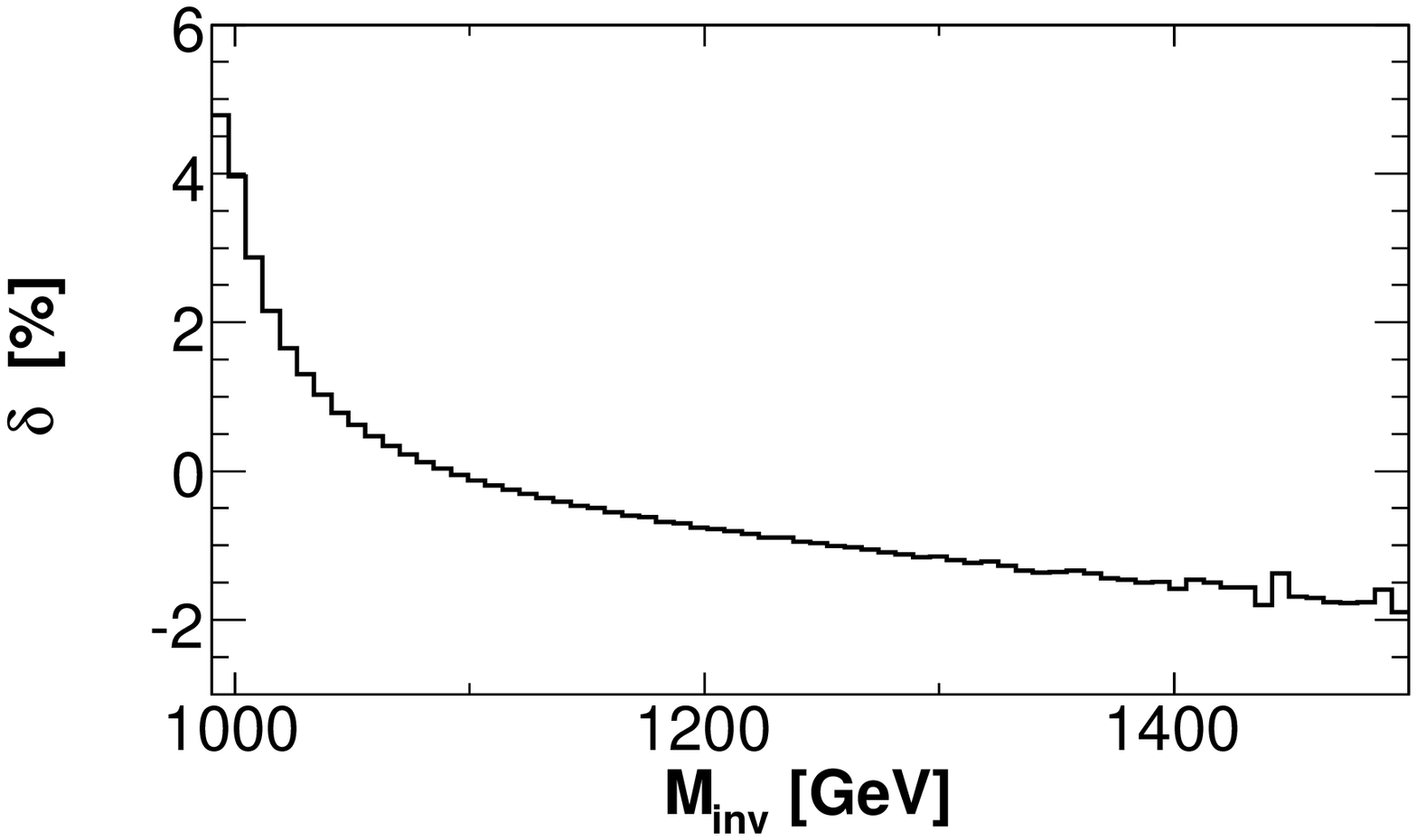, width=7cm} \\
\centering
\underline{TP2}\\
\epsfig{file=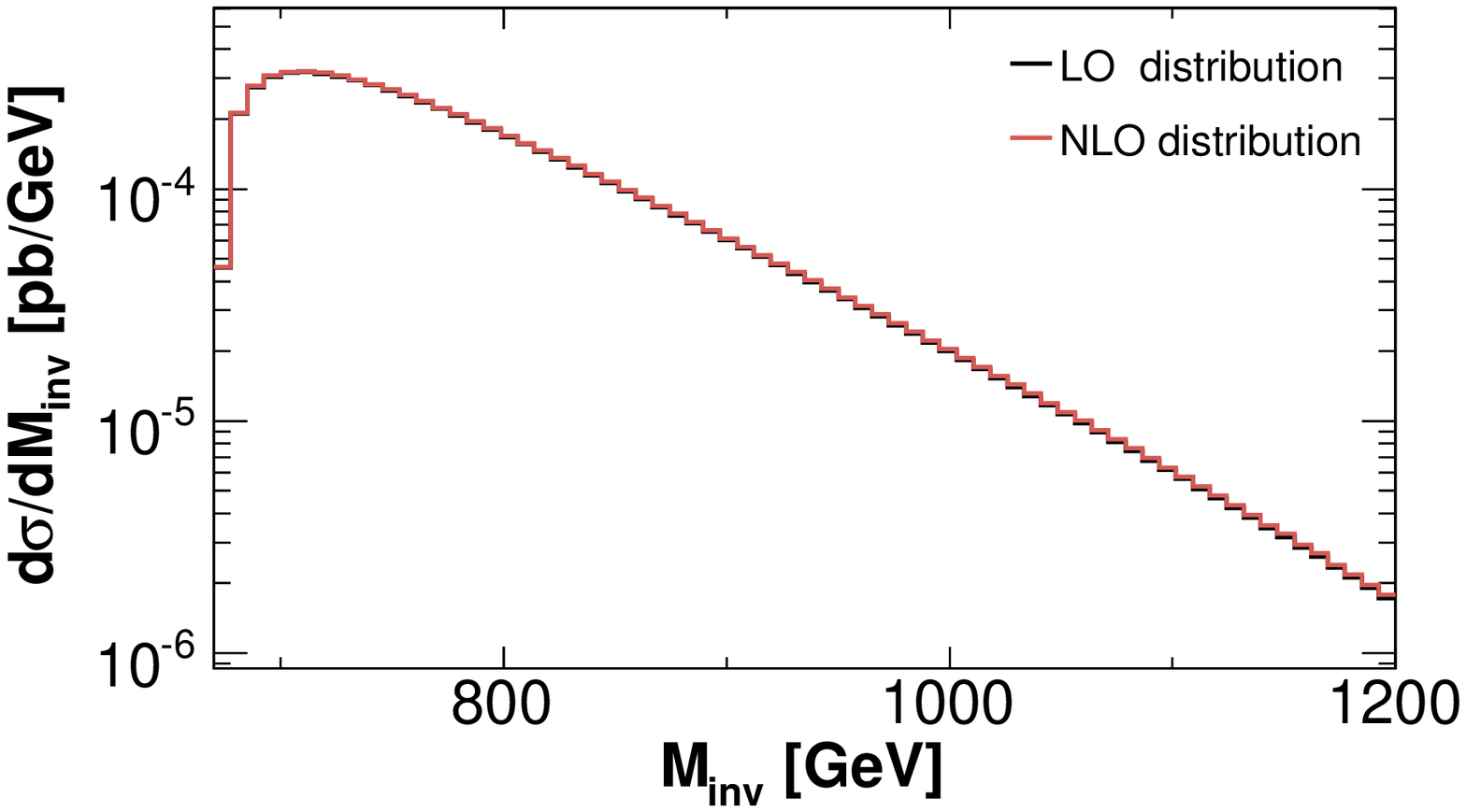, width=7cm}
\epsfig{file=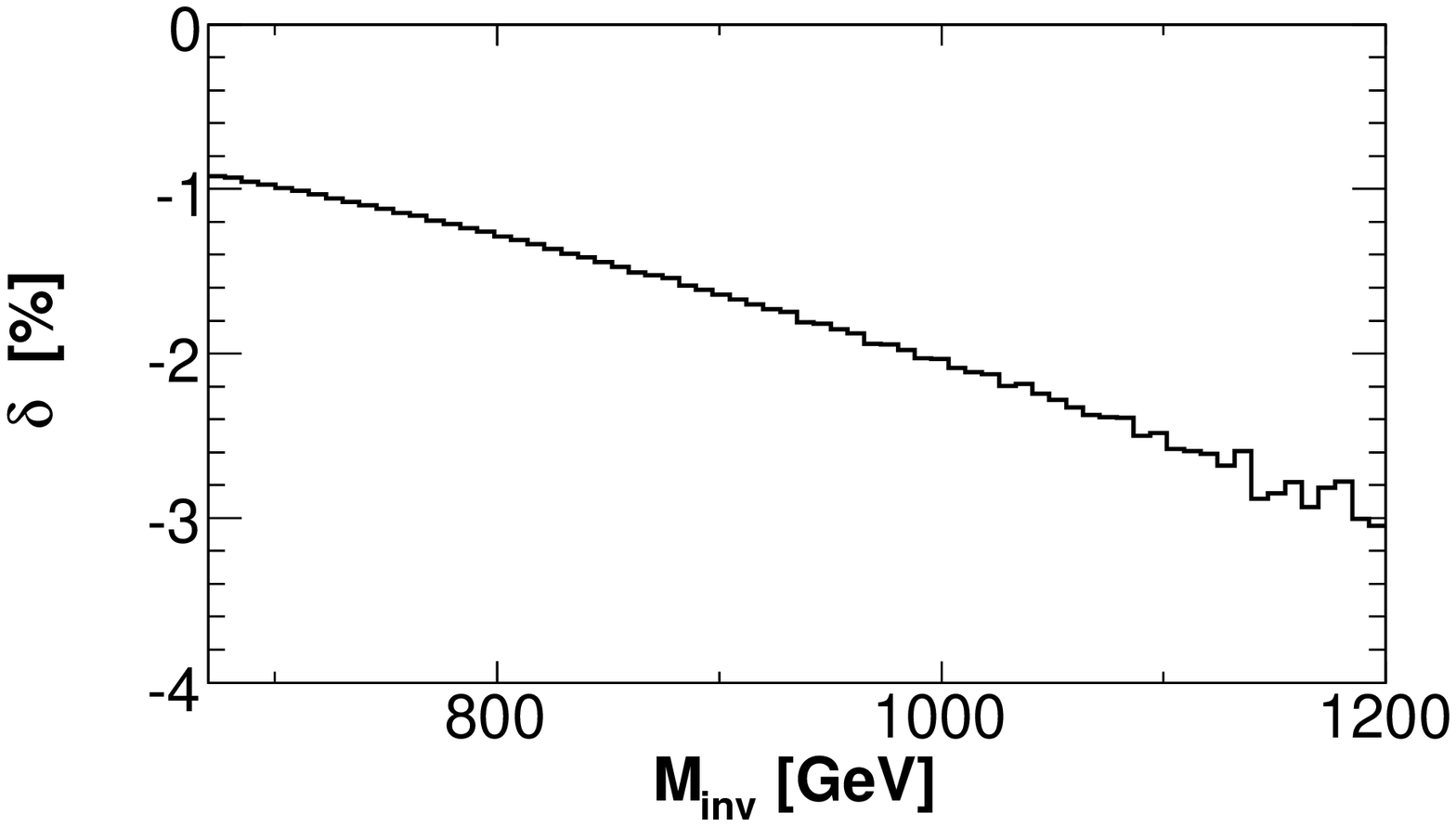, width=7cm} 
\caption{Invariant mass distribution of the two gluinos produced at
  the Tevatron via the process $P \overline{P} \to \tilde{g} \tilde{g}
  X$.  In the left panels 
we show the LO  and  the EW+NLO EW  contribution, while the electroweak corrections relative to  
the total result are shown in the right  panels.}
\label{Fig:IMtev}
\end{figure}

The invariant mass distribution for the two points is shown in
Fig.~\ref{Fig:IMtev}. In both cases
the $\mathcal{O}(\alpha_s^2\alpha)$
corrections are small compared to the lowest order results and do not change
the shape of the distribution. 
In particular,
in the case of the point TP1 (TP2)   EW corrections
relative to  the total contribution are of the order of $-2$ to $5 \%$ ($-3$ to $-1 \%$). 
\medskip

\begin{figure} 
\centering
\underline{TP1}\\
\epsfig{file=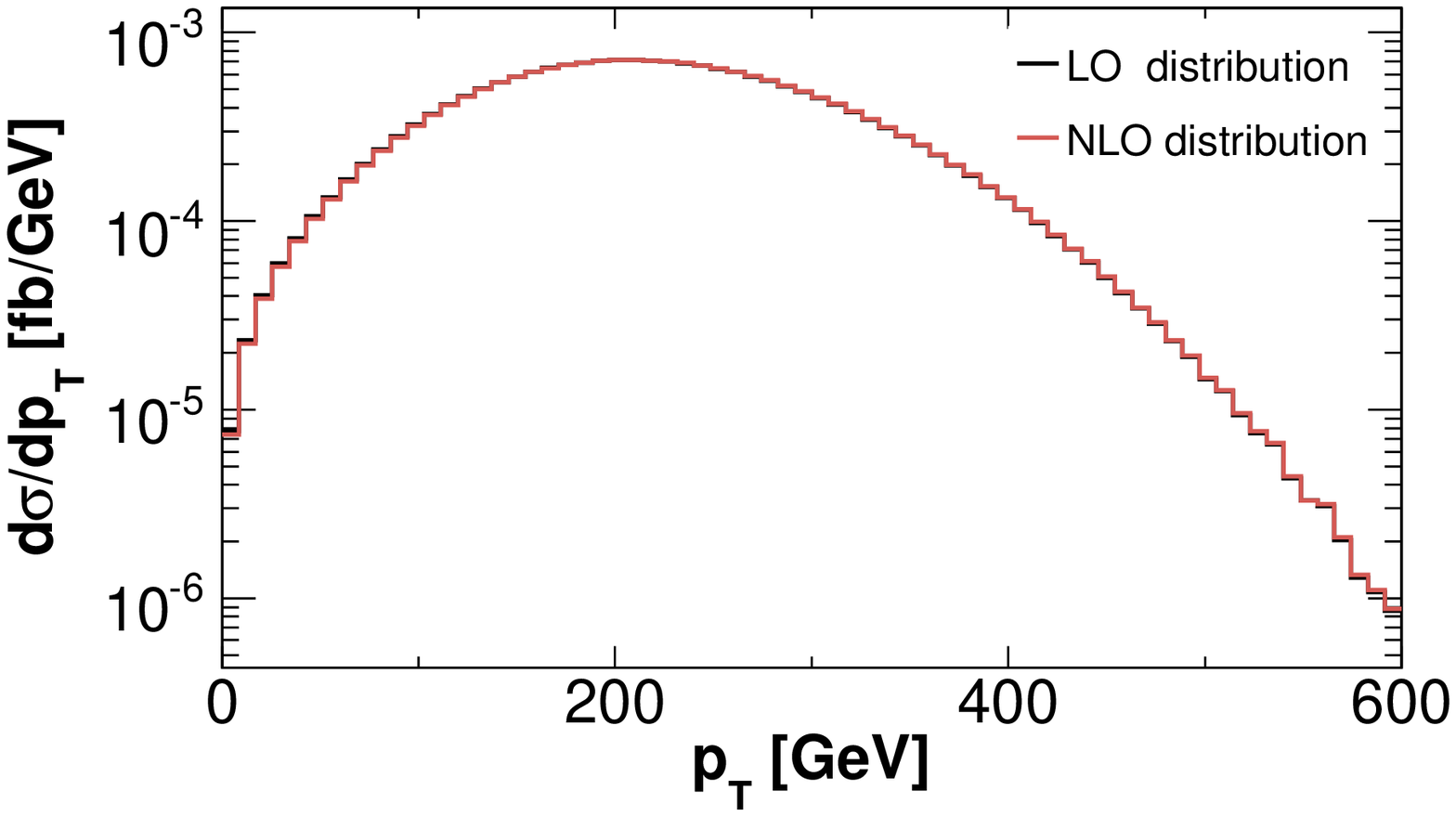, width=7cm}
\epsfig{file=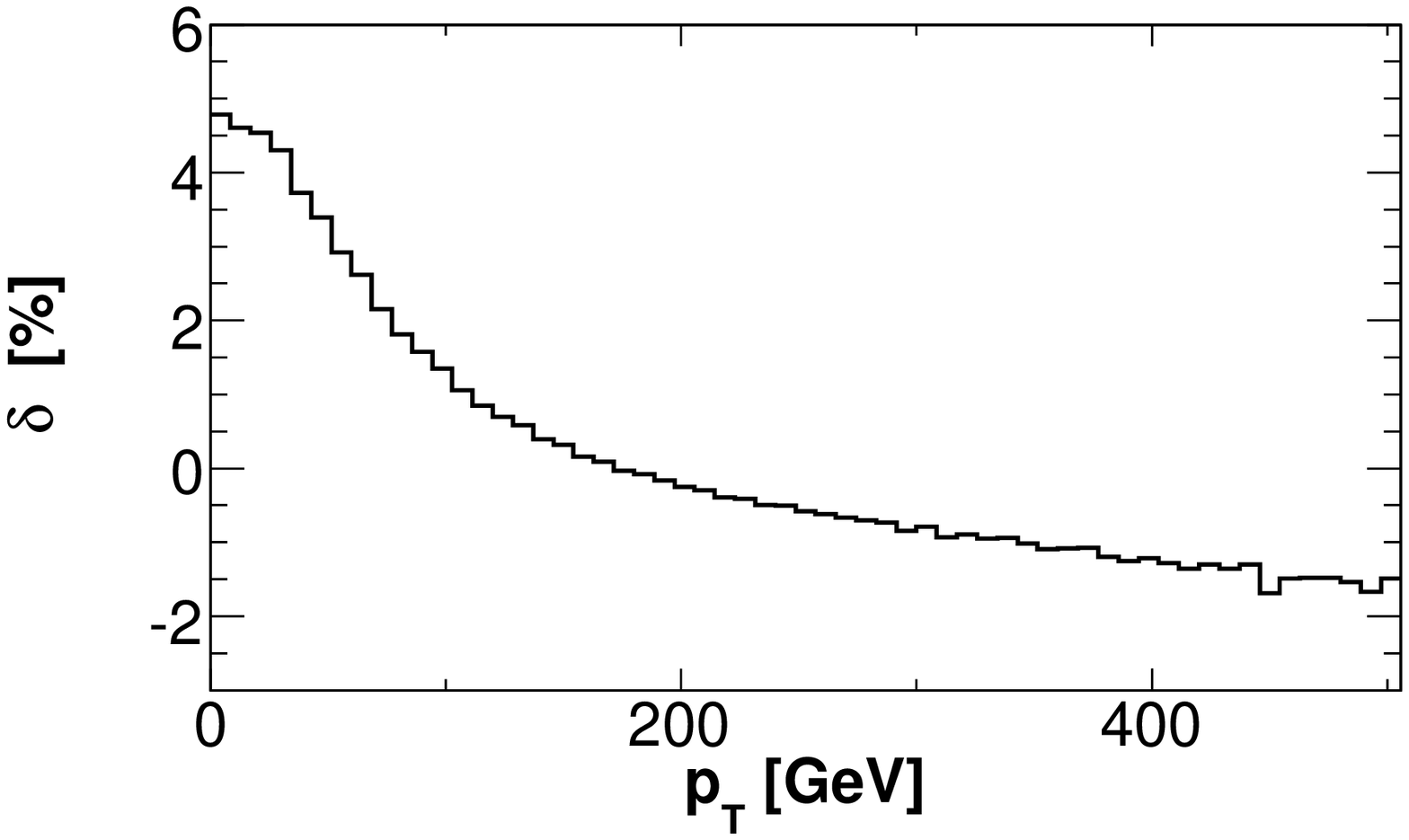, width=7cm} \\
\centering
\underline{TP2} \\
\epsfig{file=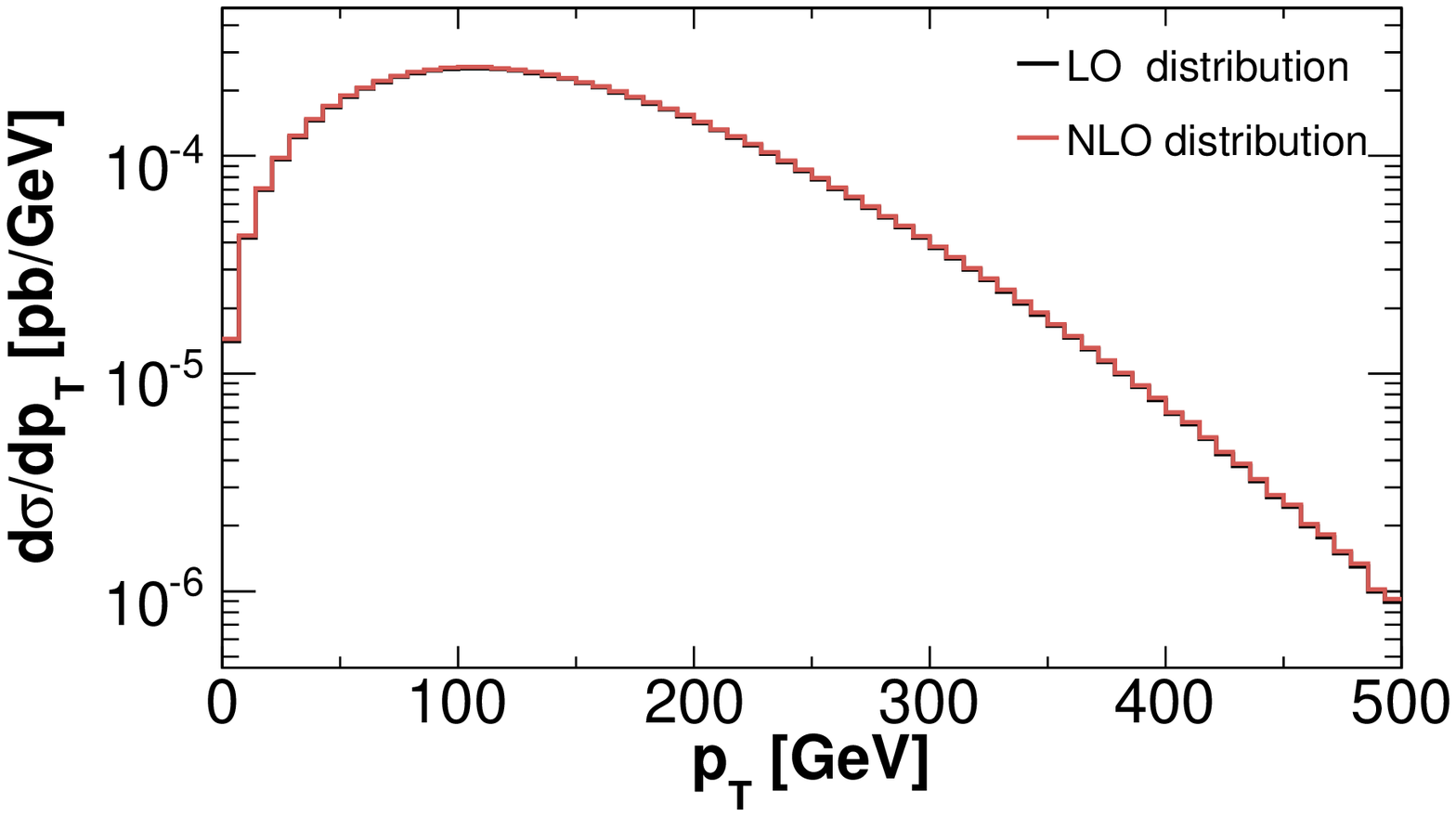, width=7cm}
\epsfig{file=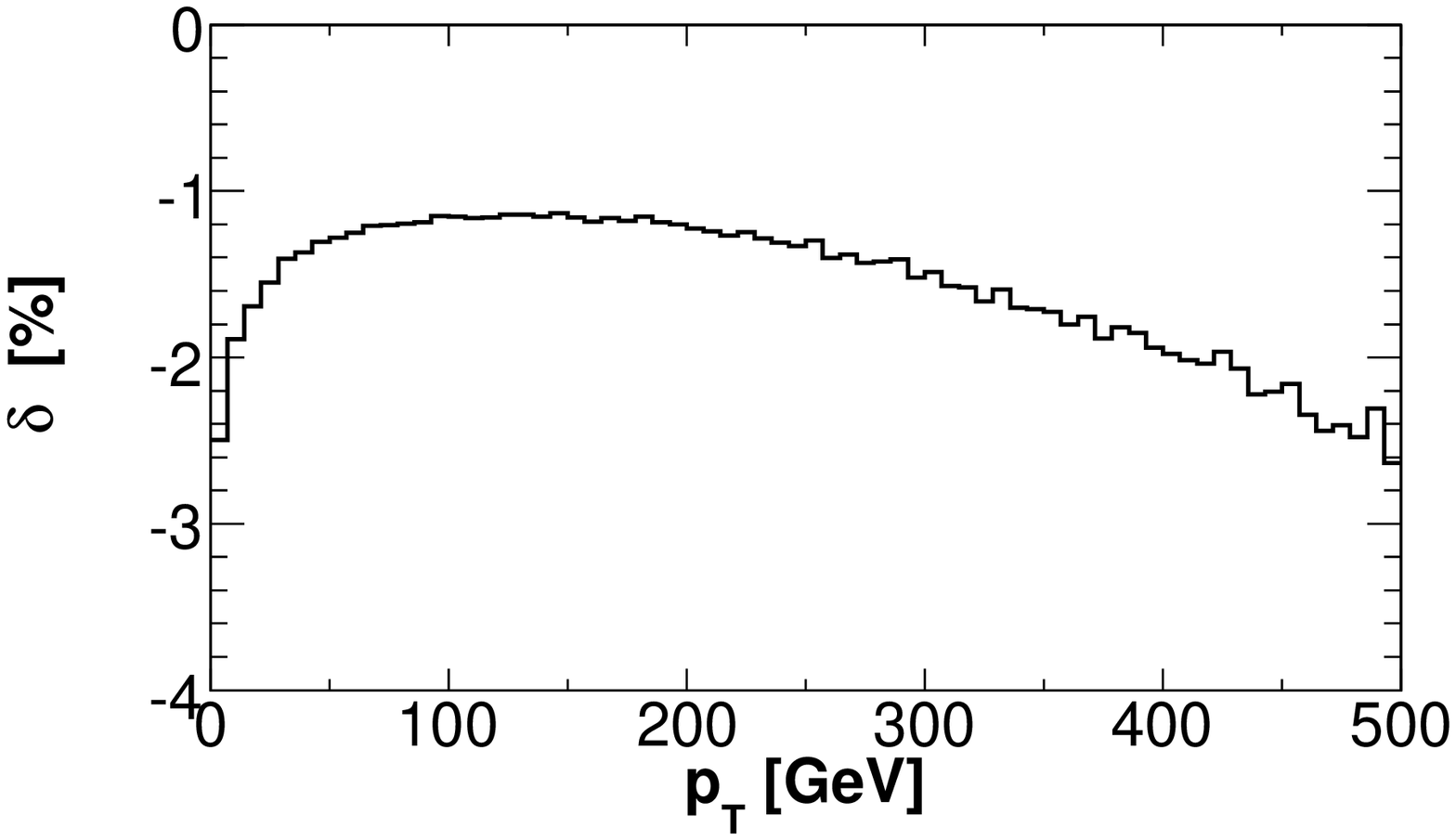, width=7cm}
\caption{Same as Fig.~\ref{Fig:IMtev}, but considering the transverse momentum distribution.}
\label{Fig:PTtev}
\end{figure}

Similar considerations hold in the case of the transverse momentum
distribution, shown in Fig.~\ref{Fig:PTtev}. The shape of the
distribution is not affected by the insertion of the electroweak corrections in both points. The
electroweak corrections  relative to the total contribution  
are  of order of $-2$ to $4.5 \%$ in the case of the TP1 point
and of the order of $-2.5$ to $-1 \%$ in the TP2 scenario.

\section{Conclusions}
\label{Conclusions}
In this paper, we have computed the full $\mathcal{O}(\alpha_s^2\alpha)$ corrections to  
gluino pair production at the LHC and at the Tevatron. Two
different renormalization schemes were used. The numerical value of the
$\mathcal{O}(\alpha_s^2\alpha)$ contribution 
is rather independent on the renormalization scheme.
The treatment of the 
IR and collinear singularities was performed within 
two different methods.
\medskip

We have studied the numerical impact of the
$\mathcal{O}(\alpha_s^2\alpha)$ contributions at the LHC in 
two different scenarios and we have performed scans over many regions
of the parameter space. The EW
corrections are negative and can be safely neglected. 
Compared to squark--anti-squark~\cite{Hollik:2007wf, Hollik:2008yi}  and squark-gluino~\cite{Hollik:2008vm} production, 
the EW contributions to gluino pair production are less important. The  main reason is that the EW contributions do not enter the gluon fusion 
channel, which is the leading tree-level production  channel in a wide part of the region of the 
parameter space investigated in this paper. 
\medskip

We have also provided numerical results for  gluino pair production at the Tevatron selecting two scenarios 
belonging to the region of the parameter space investigated  by the D0  and CDF collaborations. Again, the
$\mathcal{O}(\alpha_s^2\alpha)$ contributions are small and negligible.

\subsection*{Acknowledgments} 
We are indebted to Jan~Germer, Wolfgang~Hollik, and  Maike~Trenkel  for useful discussions and
for reading the manuscript. Michael~Rauch is gratefully
acknowledged for countless suggestions in the early stages of this work.

\clearpage


\section*{Appendix}
\appendix
%
%
%
%
%
\section{Feynman Diagrams}
\label{SSec:FeynmanDiagramsGLUGLU}
In this appendix we collect the relevant Feynman diagrams. In the following the label $S$ ($S^0$)
is used to denote charged (neutral) Higgs bosons. Moreover  $V^0=\gamma, Z$.
%
%
\begin{figure}[h]
\centering
\input{DIAG/tree}
\caption{Tree-level diagrams for the processes $q\overline{q}\to \tilde{g}\tilde{g}$ and
$gg \to \tilde{g}\tilde{g}$.}
\label{Fig:TREE}
\end{figure}
%
%
\begin{figure}[h]
\input{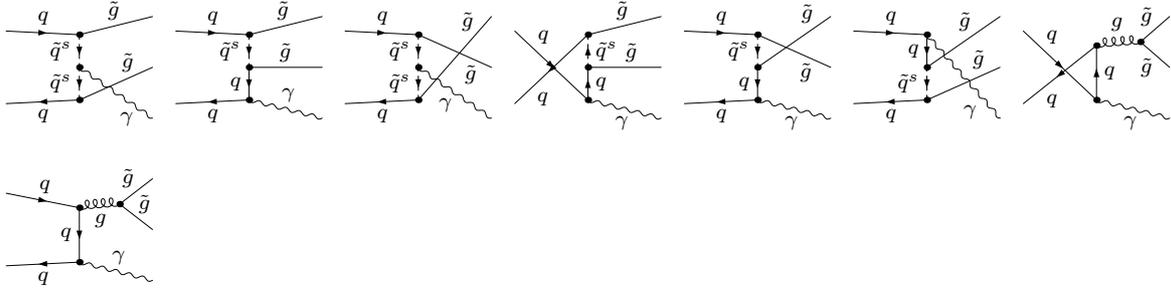}
\caption{Tree-level diagrams for the real photon emission process $q\overline{q} \to \tilde{g}\tilde{g}\gamma$. }
\label{Fig:2in3gamma}
\end{figure}
%
%
\begin{figure}[h]
\input{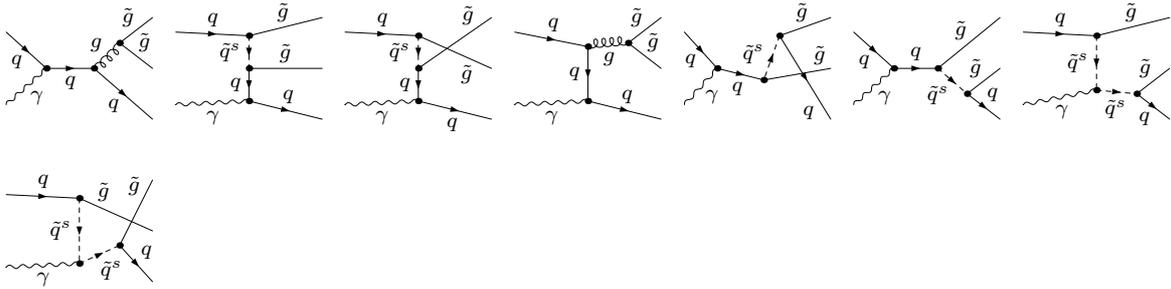}
\caption{Tree-level diagrams for the process $ q\gamma  \to \tilde{g} \tilde{g}  q$. The diagrams for the 
process $\gamma \overline{q} \to \tilde{g} \tilde{g} \overline{q}$ can be obtained inverting the arrows. }
\label{Fig:QuarkGamma}
\end{figure}
%
\begin{figure}[h]
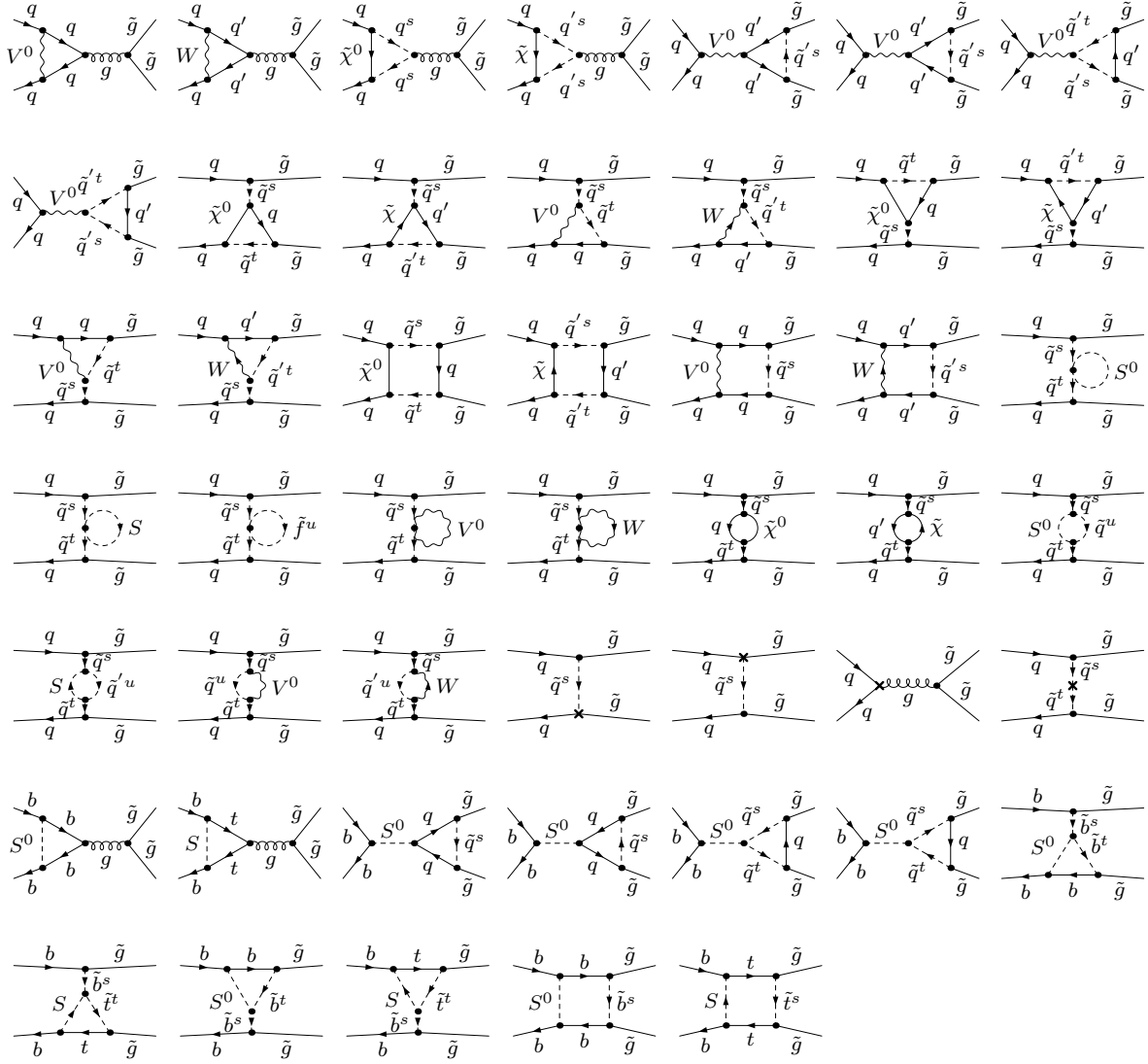
 
\input{DIAG/Loop}
\input{DIAG/LoopB}
\caption{One-loop EW diagrams for the process 
$q\overline{q} \to \tilde{g}\tilde{g}$. Diagrams with crossed final states 
are not shown.}
\label{Fig:LOOPQ}
\end{figure}
\clearpage

\newpage

\bibliographystyle{JHEP}
\bibliography{reference}

\newpage

\clearpage
\newpage

\end{document}